\documentclass{aa}

\usepackage{graphicx}
\usepackage[varg]{txfonts}
\usepackage{natbib}
\usepackage[pdftex,colorlinks=true,linkcolor=blue,citecolor=blue,urlcolor=blue]{hyperref}
\usepackage{amsmath}
\usepackage{bbold}
\usepackage{tikz}
\usepackage{bm}
\usetikzlibrary{decorations.pathreplacing,math}

\def\expo#1{\mathbf{e}^{#1}}

\def\A{\mathcal{A}}
\def\U{\mathcal{U}}
\def\V{\mathcal{V}}

\def\H{\mathcal{H}}
\def\K{\mathcal{K}}
\def\I{\mathcal{I}}
\def\J{\mathcal{J}}

\def\t{^\mathrm{T}}

\def\primet{^\mathrm{\prime T}}

\def\L{\mathcal{L}}
\def\O{\mathcal{O}}

\def\spleafURL{\url{https://gitlab.unige.ch/jean-baptiste.delisle/spleaf}}
\def\leaf{\textsc{leaf}}
\def\spleaf{\textsc{s+leaf}}
\def\spleaftwo{\textsc{s+leaf}~2}
\def\celerite{\textit{celerite}}
\def\rhk{\log R_{HK}'}

\DeclareMathOperator{\cov}{cov}
\DeclareMathOperator{\diag}{diag}
\DeclareMathOperator{\tril}{tril}
\DeclareMathOperator{\triu}{triu}

\def\hadprod{\!*\!}
\def\l#1{\left#1}
\def\r#1{\right#1}
\renewcommand*{\vec}[1]{\bm{#1}}

\graphicspath{{./figures/}}

\begin{document}

\title{Efficient modeling of correlated noise}
\subtitle{III. Scalable methods for jointly modeling several observables' time series with Gaussian processes}

\author{J.-B. Delisle\inst{1}
  \and N. Unger\inst{1}
  \and N. C. Hara\inst{1,}\thanks{NCCR CHEOPS fellow}
  \and D. Ségransan\inst{1}
}
\institute{Département d'astronomie, Université de Genève,
  chemin Pegasi 51, 1290 Versoix, Switzerland\\
  \email{jean-baptiste.delisle@unige.ch}
}

\date{\today}

\abstract{
  The radial velocity method is a very productive technique used to detect and confirm extrasolar planets.
  The most recent spectrographs, such as ESPRESSO or EXPRES,
  have the potential to detect Earth-like planets around Sun-like stars.
  However, stellar activity can induce radial velocity variations that dilute or even mimic the signature of a planet.
  A widely recognized method for disentangling these signals is to model the radial velocity time series,
  jointly with stellar activity indicators, using Gaussian processes and their derivatives.
  However, such modeling is prohibitive in terms of computational resources for large data sets,
  as the cost typically scales as the total number of measurements cubed.

  Here, we present \spleaftwo{}, a Gaussian process framework
  that can be used to jointly model several time series,
  with a computational cost that scales linearly with the data set size.
  This framework thus provides a state-of-the-art Gaussian process model,
  with tractable computations even for large data sets.

  We illustrate the power of this framework by reanalyzing the 246 HARPS radial velocity measurements of
  the nearby K2 dwarf HD~138038, together with two activity indicators.
  We reproduce the results of a previous analysis of these data,
  but with a strongly decreased computational cost (more than two order of magnitude).
  The gain would be even greater for larger data sets.
}


\maketitle

\section{Introduction}
\label{sec:intro}

It is common in astronomy to indirectly detect a physical event or the presence of a body
by searching for its signature in a data set and, more specifically, in a time series.
Astronomical time series are typically corrupted by photon noise, which is uncorrelated:
the noise values at two distinct times are statistically independent.
In that case, as more data are acquired, the searched-for signal should emerge more clearly.
However, in many cases, the data are also corrupted by
correlated noise emerging from other physical events, contamination from the Earth's atmosphere,
instrumental noise, etc.
In some cases, the structure of this correlated noise can mimic the signal of interest,
leading to spurious detections or to a poor estimation of the model parameters.

This situation is encountered in particular when searching for exoplanets in radial velocity (RV) data.
The RV of a star is the star velocity projected onto the line of sight, measured thanks to the Doppler effect.
The presence of a planetary companion induces a reflex motion of the star and thus a periodic pattern in the RV time series.
The latest generation of spectrographs,
such as ESPRESSO \citep{pepe_2021_espresso} or EXPRES \citep{blackman_2020_performance},
is able to reach a RV precision of the order of 10~cm/s,
and has the potential to discover Earth-like planets around Sun-like stars.
However, correlated noise of stellar origin complexifies this task.
The $p$-modes and granulation processes introduce correlated noise at different
timescales \citep{dumusque_2011_planetary,dumusque_2012_earthmass}.
Furthermore, the random appearance of spots and faculae at the surface of the star,
combined with the star's rotation,
introduces complex structure in the data,
which might be difficult to disentangle from low-mass planets.
At longer timescales (from hundreds to thousands of days),
the stellar magnetic cycle also induces RV variations,
as well as variations in activity indicators such as the flux in the calcium II H \& K emission lines
\citep[$\log R'_{HK}$,][]{noyes_1984_rotation} as in, for instance, \citep{queloz_2001_planet}.
The RV signals induced by these multiple physical processes of stellar origin are globally referred to as stellar activity.
A common way to account for stellar activity is to model it
through a Gaussian process (GP) model \citep{rasmussen_2006_gaussian}.
The GP regression method allows for the modeling of complex processes by parametrizing the covariance
between the measurements instead of defining a deterministic model of the physical processes.
For a GP $G(t)$ measured at times $t_i$ and $t_j$,
the values $G(t_i)$ and $G(t_j)$ are assumed to be randomly drawn from a normal distribution,
with covariance $C_{i,j} = k(t_i, t_j)$,
where $k$ is the chosen parametrized kernel.
The GP is often assumed to be stationary, such that the kernel $k$ only depends
on the lag $\Delta t = |t_i-t_j|$ between two measurements.
A commonly used kernel to model stellar activity is \citep{aigrain_2012_simple,haywood_2014_planets,rajpaul_2015_gaussian}:
\begin{equation}
  \label{eq:SEPkernel}
  k(\Delta t) = \sigma^2 \exp \l(- \frac{\Delta t^2}{2 \rho^2}
  - \frac{\sin^2 \l( \frac{\pi \Delta t }{P}\r) }{2 \eta^2}\r),
\end{equation}
where $\sigma$, $P$, $\rho$, and $\eta$ are the hyperparameters of the GP which need to be adjusted.
In the following, we refer to this kernel as the squared-exponential periodic (SEP) kernel.

Gaussian processes are known to be able to represent a wide range of signals.
As such, when their hyperparameters are left free,
they are prone to absorb the signal of interest (planetary signal)
along with the correlated noise (stellar activity).
To avoid this drawback,
\citet{rajpaul_2015_gaussian} proposed a framework
in which the RV time series is modeled jointly with activity indicators.
Building on \citep{aigrain_2012_simple},
the authors assume that the activity-induced variations of the measurements
depend linearly on an underlying Gaussian process $G(t)$
and its derivative $G'(t)$.
The evolution of the RV and indicators is modeled as:
\begin{align}
  \label{eq:RajpaulModel}
   & \Delta \mathrm{RV} = V_c G(t) + V_r G'(t),\nonumber  \\
   & \Delta \mathrm{BIS} = B_c G(t) + B_r G'(t),\nonumber \\
   & \Delta \rhk = L_c G(t),
\end{align}
for some constants $V_c,V_r,B_c,B_r,L_c$.
The GP's hyperparameters are thus constrained by the three time series, instead of the RV alone.
This reduces the risk of the GP overfitting, that is, the absorbtion of planetary signals,
since those signals are only present in the RV time series.
This framework can be straightforwardly generalized to account for additional indicators,
for the combination of several GP with different amplitudes,
or even for the second order derivatives of the GP \citep[e.g.,][]{jones_2017_improving}.

While this framework is very powerful in modeling stellar activity,
it represents a challenge in terms of computational cost.
Indeed, computing the likelihood (or $\chi^2$) of the model for a given set of hyperparameters
requires us to solve a linear system involving the full covariance matrix of the measurements.
For a time series of size $n$, the full covariance matrix
-- including RV, BIS, and $\rhk$ measurements --
has a size of $3n \times 3n$,
and the cost of the solving typically scales as $\O\l((3n)^3\r)$.
This becomes prohibitive in terms of computer time for large data sets,
especially in the context of Bayesian methods (MCMC, nested sampling, etc.),
which might require billions of evaluations of the likelihood.

In the context of a GP applied to a single time series,
\citet{ambikasaran_2015_generalized} and \citet{foreman-mackey_2018_scalable}
\citep[see also][]{rybicki_1995_class}
have shown that the so-called \celerite{} kernel,
\begin{equation}
  \label{eq:celerite}
  k(\Delta t) = \sum_{s<n_\mathrm{c}}
  \l(a_s \cos(\nu_s\Delta t) + b_s \sin(\nu_s\Delta t) \r) \expo{-\lambda_s\Delta t},
\end{equation}
where $n_c$ is the number of components, and $a_s$, $b_s$, $\lambda_s$, and $\nu_s$ are
the kernel hyperparameters,
can be represented as a semiseparable matrix.
As a consequence, the computational cost of evaluating the likelihood
scales linearly with the number of points ($\O\l(n\r)$),
allowing to apply these methods to large data sets.
\citet{delisle_2020_efficientb} defined a more general class of covariance matrices
with a similar linear scaling of the cost:
the \spleaf{} matrix, which is the sum of a semiseparable matrix and a \leaf{} matrix.
The \leaf{} component, which has non-zero elements close to the diagonal,
is particularly adapted to represent calibration noise \citep[see][]{delisle_2020_efficientb}.
\citet{gordon_2020_fast} extended the \celerite{} model to the case of two-dimensional data sets.
This applies, in particular, to the case of several parallel time series (e.g., RV, BIS, and $\rhk$),
with measurements taken at the same times.
However, \citet{gordon_2020_fast} do not discuss the treatment of the derivatives of the GP,
and thus of models similar to the \citet{rajpaul_2015_gaussian} model
(Eq.~(\ref{eq:RajpaulModel})).

In this study, we extend the \celerite{} and \spleaf{} models
to account for the case of several time series, with independent calendars,
modeled as a linear combination of several GP and their derivatives.
This allows us to apply models similar to the model used by \citet{rajpaul_2015_gaussian},
but with a linear scaling of the evaluation cost of the likelihood.
We call this new model \spleaftwo{},
as it is a generalization of the \spleaf{} model \citep{delisle_2020_efficientb}.

In Sect.~\ref{sec:homogeneous}, we recall the main properties of the \celerite{} and \spleaf{} models.
We then show in Sect.~\ref{sec:derivative} how to model the derivative of a \celerite{} GP.
In Sect.~\ref{sec:heterogeneous}, we extend the model to the case of multiple time series.
We illustrate the power of this framework by reanalyzing the HARPS RV
of the nearby K2 dwarf \object{HD~13808} in Sect.~\ref{sec:application}.
Finally, we discuss our methods and results in Sect.~\ref{sec:conclusion}.
An open-source reference implementation of our algorithms as C library
with python wrappers is publicly available
\footnote{\spleafURL}.

\section{The \celerite{} and \spleaf{} models for homogeneous time series}
\label{sec:homogeneous}

We consider a time series of measurements $(t_i, y_i)$ ($i=1,\dots,n$),
which can be modeled by a deterministic component, a GP component, and measurement noise.
In the case of radial velocities,
the deterministic component encompasses the reflex motion due to companions,
the systematic velocity of the system,
instruments offsets, and so on.
The GP might be used to model physical mechanisms that are  too poorly understood or constrained
to be included in the deterministic part.
This is typically the case of stellar activity at different timescales
(oscillations, granulation, rotation, magnetic cycles, etc.).
Finally the noise component encompasses photon noise, calibration noise, and so on.
The time series can thus be expressed as:
\begin{equation}
  y_i = m(t_i) + G(t_i) + \epsilon_i,
\end{equation}
where
$m$ is the deterministic part of the model, $G$ is the GP, and $\epsilon$ the noise.
These three components of the model might depend on a set of parameters $\theta$.
Assuming the noise to also be Gaussian (but not necessarily white),
the log-likelihood function of a given set of parameters $\theta$ is
\begin{align}
  \label{eq:loglike}
  \ln\L(\theta) & = \ln p(y|\theta)\nonumber                                                        \\
                & = -\frac{1}{2} \Big(y-m_\theta\Big)\t C_\theta^{-1} \Big(y-m_\theta\Big)\nonumber \\
                & \quad -\frac{1}{2} \ln\det \Big(2\pi C_\theta\Big).
\end{align}
$C$ is the total covariance matrix of the time series which can be split as
\begin{equation}
  C = K + \Sigma,
\end{equation}
where $K$ is the covariance of the GP $G$
and $\Sigma$ is the covariance of the noise
\begin{align}
  K_{i,j}      & = \cov\l(G(t_i), G(t_j)\r) = k(t_i, t_j),\nonumber \\
  \Sigma_{i,j} & = \cov\l(\epsilon_i, \epsilon_j\r),
\end{align}
with $k$ the GP kernel function.

\subsection{The \celerite{} model}
\label{sec:celerite1}

The \celerite{} model proposed by \citet{foreman-mackey_2017_fast}
allows for very efficient computations of this model
and, in particular, the likelihood and its gradient with respect to $\theta$,
in the case of white noise (diagonal covariance matrix $\Sigma = \diag(\sigma^2)$)
and assuming the kernel function $k$ to follow Eq.~(\ref{eq:celerite}).
In this case, the covariance matrix $C$
is semiseparable \citep[see][]{foreman-mackey_2017_fast}:
\begin{equation}
  \label{eq:semisep}
  C = \diag\l(A+\sigma^2\r) + \tril\l(U V\t\r) + \triu\l(V U\t\r),
\end{equation}
where $\diag(A)$ is the diagonal matrix built from the vector $A$ of size $n$,
$U$ and $V$ are $n\times r$ matrices (with $r=2n_c$ the rank of $U$ and $V$)
defined as:
\begin{align}
  \label{eq:celeritedecomp}
  A_i                  & = \sum_{s<n_\mathrm{c}}a_s,\nonumber                                              \\
  U_{i,s}              & = \expo{-\lambda_s t_i} \l(a_s \cos(\nu_s t_i) + b_s \sin(\nu_s t_i)\r),\nonumber \\
  U_{i,n_\mathrm{c}+s} & = \expo{-\lambda_s t_i} \l(a_s \sin(\nu_s t_i) - b_s\cos(\nu_s t_i)\r),\nonumber  \\
  V_{i,s}              & = \expo{\lambda_s t_i} \cos(\nu_s t_i),\nonumber                                  \\
  V_{i,n_\mathrm{c}+s} & = \expo{\lambda_s t_i} \sin(\nu_s t_i).
\end{align}
This low-rank representation of the covariance matrix allows us
to use very efficient dedicated algorithms
\citep[Cholesky decomposition, solving, dot product, determinant, see][]{foreman-mackey_2017_fast,foreman-mackey_2018_scalable},
which are particularly useful in computing the likelihood.
The memory footprint of the \celerite{} model scales as $\O\l(nr\r)$,
and the computational cost scales as $\O\l(nr^2\r)$.
For comparison, the memory footprint of a naive representation of the same covariance matrix
scales as $\O\l(n^2\r)$ and the computational costs typically scales as $\O\l(n^3\r)$.

The \celerite{} algorithm is actually not restricted to kernels of the form of
Eq.~(\ref{eq:celerite}), but it can be applied as long as the kernel admits a semiseparable
representation following Eq.~(\ref{eq:semisep}).
In particular, the Matérn 3/2 and Matérn 5/2 can be represented using a semiseparable representation
of rank 2 and 3 respectively (see Appendix~\ref{sec:matern_semisep}).

\subsection{The \spleaf{} model}
\label{sec:spleaf1}

The \spleaf{} model \citep{delisle_2020_efficientb}
extends the \celerite{} model in the case of correlated measurement noise.
While in the \celerite{} model, the covariance matrix $\Sigma$
is assumed to be diagonal,
the \spleaf{} model allows us to account for a more general class of
noise models, where $\Sigma$ can be represented by a \leaf{} matrix.
Such \leaf{} matrices are sparse matrices where non-zero values are close to the diagonal
\citep[see][]{delisle_2020_efficientb}.
This encompasses banded matrices, block-diagonal matrices, staircase matrices, and so on.
In the context of radial velocities, \leaf{} matrices are especially useful to model
calibration noise \citep[see][]{delisle_2020_efficientb}.
Indeed, to obtain precise RV measurements,
the instrument must be calibrated periodically, typically once per night.
Thus, several measurements taken with the same instrument during the same night share the same calibration noise,
which introduces a block-diagonal contribution to the covariance matrix.

As in the case of the \celerite{} model,
the \spleaf{} model allows for a sparse representation of the covariance matrix
and for efficient dedicated algorithms
to compute the likelihood and its gradient with respect to the parameters \citep{delisle_2020_efficientb}.
The memory footprint of the \spleaf{} model scales as $\O\l(\l(r+\bar{b}\r)n\r)$
and its computational cost as $\O\l(\l(r^2+r\bar{b}+\bar{b}^2\r)n\r)$,
where $\bar{b}$ is the average band width of the $\leaf{}$ component.

\section{Derivative of a \celerite{}/\spleaf{} Gaussian process}
\label{sec:derivative}

We consider a GP $G(t)$ whose kernel function $k$ is stationary and admits a semiseparable decomposition (Eq.~(\ref{eq:semisep})).
The covariance matrix of $G$ is thus given by:
\begin{equation}
  \cov\l(G(t_i), G(t_j)\r) = k(t_i - t_j) = U(t_i) V(t_j) = U_i V_j,
\end{equation}
for $t_i \geq t_j$, namely, for the lower triangular part.
By way of symmetry, we have
\begin{equation}
  \cov\l(G(t_i), G(t_j)\r) = k(t_j - t_i) = V_i U_j,
\end{equation}
for $t_i < t_j$, namely, for the upper triangular part.
Assuming $G$ to be differentiable, the covariance between $G$ and $G'$
is given by the partial derivatives of the kernel function:
\begin{align}
  \label{eq:covGdG}
  \cov\l(G'(t_i), G(t_j)\r) & = \frac{\partial k(t_i, t_j)}{\partial t_i} = k'(t_i-t_j)
  = U'(t_i) V(t_j) = U'_i V_j,\nonumber                                                  \\
  \cov\l(G(t_i), G'(t_j)\r) & = \frac{\partial k(t_i, t_j)}{\partial t_j} = -k'(t_i-t_j)
  = U_i V'_j,
\end{align}
where we assume $t_i \geq t_j$,
and the primes denote the differentiation with respect to time.
For these two equations to be valid,
the semiseparable representation $(U,V)$ must verify:
\begin{equation}
  \label{eq:stationarity}
  U' V\t = - U V\primet.
\end{equation}
This relation is actually linked with the stationarity of the kernel.
Indeed, since
\begin{equation}
  k(t, t+\Delta t) = k(\Delta t) = U(t) V(t+\Delta t)
\end{equation}
does not depend on $t$,
we have
\begin{equation}
  \frac{\partial k(t, t+\Delta t)}{\partial t} = 0
  = U'(t) V(t+\Delta t) + U(t) V'(t+\Delta t).
\end{equation}
In addition to this stationarity condition, for the GP $G$ to be differentiable
the kernel must verify $k'(0) = 0$ \citep[e.g.,][]{rasmussen_2006_gaussian},
thus:
\begin{equation}
  \label{eq:differentiability}
  U'_i V_i = U_i V'_i = 0.
\end{equation}
From Eq.~(\ref{eq:covGdG}),
we deduce that
the covariance matrix between $G$ and $G'$ admits
several equivalent antisymmetric semiseparable representations
\begin{align}
  \label{eq:Dantisep}
  \cov\l(G', G\r) & = \tril\l(U' V\t\r) - \triu\l(V {U'}\t\r)\nonumber   \\
                  & = -\tril\l(U {V'}\t\r) + \triu\l(V' U\t\r),\nonumber \\
                  & = \tril\l(U' V\t\r) + \triu\l(V' U\t\r),\nonumber    \\
                  & \dots
\end{align}
Similar representations can be deduced for $\cov\l(G, G'\r)$
by using the relation
\begin{equation}
  \cov\l(G, G'\r) = \cov\l(G', G\r)\t = - \cov\l(G', G\r).
\end{equation}
For the covariance matrix of $G'$ itself, we have (for $t_i \geq t_j$)
\begin{align}
  \label{eq:covdGdG}
  \cov\l(G'(t_i), G'(t_j)\r) & = \frac{\partial^2 k(t_i, t_j)}{\partial t_i\partial t_j} = -k''(t_i-t_j)\nonumber \\
                             & = U'_i V'_j = -U''_i V_j = -U_i V''_j.
\end{align}
Therefore, the covariance matrix of $G'$
also admits several symmetric semiseparable representations:
\begin{align}
  \label{eq:D2semisep}
  \cov\l(G', G'\r) & = \diag(B) + \tril\l(U' {V'}\t\r) + \triu\l(V' {U'}\t\r)\nonumber \\
                   & = \diag(B) - \tril\l(U'' V\t\r) - \triu\l(V {U''}\t\r)\nonumber   \\
                   & = \diag(B) - \tril\l(U {V''}\t\r) - \triu\l(V'' U\t\r)\nonumber   \\
                   & \dots
\end{align}
with
\begin{equation}
  B_i = U'_i V'_i = -U''_i V_i = - U_i V''_i.
\end{equation}
Hereinafter, we use the following representations:
\begin{align}
  \label{eq:DD2}
  \cov\l(G', G\r)  & = \tril\l(U' V\t\r) + \triu\l(V' U\t\r),\nonumber         \\
  \cov\l(G, G'\r)  & = \tril\l(U {V'}\t\r) + \triu\l(V {U'}\t\r),\nonumber     \\
  \cov\l(G', G'\r) & = \diag(B) + \tril\l(U' {V'}\t\r) + \triu\l(V' {U'}\t\r).
\end{align}

Applying this reasoning to the \celerite{} kernel of Eq.~(\ref{eq:celerite}),
we find (as per Eq.~(\ref{eq:celeritedecomp})):
\begin{align}
  \label{eq:Dceleritedecomp}
  k'(\Delta t)          & = \sum_{s<n_\mathrm{c}}
  \l(a'_s \cos(\nu_s\Delta t) + b'_s \sin(\nu_s\Delta t) \r) \expo{-\lambda_s\Delta t},\nonumber                   \\
  -k''(\Delta t)        & = \sum_{s<n_\mathrm{c}}
  \l(a''_s \cos(\nu_s\Delta t) + b''_s \sin(\nu_s\Delta t) \r) \expo{-\lambda_s\Delta t},\nonumber                 \\
  U'_{i,s}              & = \expo{-\lambda_s t_i} \l(a'_s \cos(\nu_s t_i) + b'_s \sin(\nu_s t_i)\r),\nonumber      \\
  U'_{i,n_\mathrm{c}+s} & = \expo{-\lambda_s t_i} \l(a'_s \sin(\nu_s t_i) - b'_s\cos(\nu_s t_i)\r),\nonumber       \\
  V'_{i,s}              & = \expo{\lambda_s t_i} \l(\lambda_s \cos(\nu_s t_i) - \nu_s \sin(\nu_s t_i)\r),\nonumber \\
  V'_{i,n_\mathrm{c}+s} & = \expo{\lambda_s t_i} \l(\lambda_s \sin(\nu_s t_i) + \nu_s \cos(\nu_s t_i)\r),
\end{align}
with
\begin{align}
  a'_s  & = \nu_s b_s-\lambda_s a_s,\nonumber                         \\
  b'_s  & = -\nu_s a_s-\lambda_s b_s,\nonumber                        \\
  a''_s & = (\nu_s^2-\lambda_s^2) a_s + 2\lambda_s\nu_s b_s,\nonumber \\
  b''_s & = (\nu_s^2-\lambda_s^2) b_s - 2\lambda_s\nu_s a_s.
\end{align}
In this case, the differentiability condition of Eq.~(\ref{eq:differentiability}) can be rewritten as
\begin{equation}
  k'(0) = 0 = \sum_{s<n_\mathrm{c}} a'_s.
\end{equation}
Thus the initial parameters must verify
\begin{equation}
  \sum_{s<n_\mathrm{c}} \nu_s b_s-\lambda_s a_s = 0,
\end{equation}
for the GP to be differentiable.
In particular, in the case of a kernel including
a single \celerite{} component ($n_c=1$),
one must verify $a'_s = \nu_s b_s-\lambda_s a_s = 0$ and
Eq.~(\ref{eq:Dceleritedecomp}) is simplified into
\begin{align}
  k'(\Delta t)          & = b'_s \sin(\nu_s\Delta t) \expo{-\lambda_s\Delta t},\nonumber \\
  U'_{i,s}              & = b'_s \expo{-\lambda_s t_i} \sin(\nu_s t_i),\nonumber         \\
  U'_{i,n_\mathrm{c}+s} & = -b'_s \expo{-\lambda_s t_i} \cos(\nu_s t_i).
\end{align}
The SHO kernel, which is a particular type of \celerite{} kernel
that only depends on three free parameters,
always verifies this differentiability condition.
It thus provides a smoother GP than the general \celerite{} kernel,
which is why \citet{foreman-mackey_2017_fast} recommended its application.

In the case of the Matérn 3/2 and 5/2 kernels,
similar semiseparable decompositions can be achieved
for the derivatives (see Appendix~\ref{sec:matern_deriv}).
More generally, Eq.~(\ref{eq:DD2}) provides the semiseparable decomposition
of the derivatives for any differentiable semiseparable kernel.

\section{\spleaftwo{}: Extending \celerite{}/\spleaf{} to heterogeneous time series}
\label{sec:heterogeneous}

Following \citet{rajpaul_2015_gaussian}, we assume
that the time series of the radial velocities and the different indicators follow
\begin{equation}
  \label{eq:RajpaulGeneral}
  Y_{i,j} = f_i(T_{i,j}) + \sum_k \l(\alpha_{k,i} G_k(T_{i,j}) + \beta_{k,i} G'_k(T_{i,j})\r) + \epsilon_{i,j},
\end{equation}
where $(T_{1,.}, Y_{1,.})$ is the RV time series
and $(T_{i,.}, Y_{i,.})$ are the indicators time series ($i>1$),
$f_i$ is the determinist part of the model for the time series $i$,
$G_k$ are independent GP,
and $\epsilon$ is the measurement noise (including photon noise, calibration noise, etc.).

The times and number of measurements need not be the same for all time series,
which implies that $T$ and $Y$ are not necessarily matrices
but collections of vectors of variable length.
In the case of the model presented in Eq.~(\ref{eq:RajpaulModel}),
the activity indicators (BIS and $\rhk$) are typically extracted from the
same spectra as the RV time series and, thus, they share the same sampling.
However, activity indicators can be extracted from other instruments or techniques.
For instance, \citet{haywood_2014_planets} trained a GP on the CoRoT light curve of CoRoT~7
to then use it to model the impact of stellar activity on the HARPS radial velocity time series of the same star.
While both data sets were roughly contemporary, the time series did not share the same sampling.
In such a case, we could define $(T_{1,.}, Y_{1,.})$ as the RV time series,
and $(T_{2,.}, Y_{2,.})$ as the photometric time series (with $T_1 \neq T_2$),
and model both time series jointly according to Eq.~(\ref{eq:RajpaulGeneral}).

For the sake of readability, we consider in the following a single GP $G$, such that
\begin{equation}
  \label{eq:RajpaulSingle}
  Y_{i,j} = f_i(T_{i,j}) + \alpha_i G(T_{i,j}) + \beta_i G'(T_{i,j}) + \epsilon_{i,j},
\end{equation}
but the reasoning holds in the more general case of Eq.~(\ref{eq:RajpaulGeneral}).
The covariance matrix corresponding to Eq.~(\ref{eq:RajpaulSingle}) is
\begin{align}
  \label{eq:RajpaulCov}
  \cov(Y_{i,j}, Y_{l,m}) =
   & \ \alpha_i\alpha_l \cov\l(G(T_{i,j}), G(T_{l,m})\r)\nonumber  \\
   & + \beta_i \alpha_l \cov\l(G'(T_{i,j}), G(T_{l,m})\r)\nonumber \\
   & + \alpha_i \beta_l \cov\l(G(T_{i,j}), G'(T_{l,m})\r)\nonumber \\
   & + \beta_i \beta_l \cov\l(G'(T_{i,j}), G'(T_{l,m})\r)\nonumber \\
   & + \cov\l(\epsilon_{i,j}, \epsilon_{l,k}\r).
\end{align}
The full covariance matrix is a $n\times n$ matrix, where $n$ is the total number of measurements
(radial velocities and indicators).
The cost of evaluating the corresponding likelihood - which requires to solve a linear system involving the covariance matrix
and computing its determinant - typically scales as $\O\l(n^3\r)$.
Therefore, a direct evaluation of the covariance matrix becomes rapidly prohibitive in terms of memory footprint
and computing time.

\subsection{Semiseparable representation of the model}
\label{sec:celerite2}

In order to construct a semiseparable representation of the covariance matrix of Eq.~(\ref{eq:RajpaulCov})
we merge all time vectors $T_i$ into a single vector $t$, and all data vectors $Y_i$ into a single vector $y$.
We additionally sort the measurements by increasing time;
thus, the measurements of the different time series are completely mixed in the merged time series $(t,y)$.
For a measurement $(t_k, y_k)$ we denote by $\I_k$ (index of the original time series the measurement belongs to)
and $\J_k$ (index of the measurement in this original time series)
the corresponding couple of indices of this measurement in $T$, $Y$.
The model of Eq.~(\ref{eq:RajpaulSingle}) can be rewritten as
\begin{equation}
  \label{eq:RajpaulFlat}
  y_k = f_{\I_k}(t_k) + \vec{\alpha}_k G(t_k) + \vec{\beta}_k G'(t_k) + \epsilon_{\I_k,\J_k},
\end{equation}
with $\vec{\alpha}_k = \alpha_{\I_k}$ and $\vec{\beta}_k = \beta_{\I_k}$.
The corresponding covariance matrix is (as per Eq.~(\ref{eq:RajpaulCov})):
\begin{align}
  C = \cov(y, y) =
   & \ \l(\vec{\alpha}\vec{\alpha}\t\r) \hadprod \cov\l(G(t), G(t)\r)\nonumber  \\
   & + \l(\vec{\beta} \vec{\alpha}\t\r) \hadprod \cov\l(G'(t), G(t)\r)\nonumber \\
   & + \l(\vec{\alpha} \vec{\beta}\t\r) \hadprod \cov\l(G(t), G'(t)\r)\nonumber \\
   & + \l(\vec{\beta} \vec{\beta}\t\r) \hadprod \cov\l(G'(t), G'(t)\r)\nonumber \\
   & + \Sigma,
\end{align}
where $\Sigma$ is the covariance matrix of the measurement noise
\begin{equation}
  \Sigma_{k,l} = \cov(\epsilon_{\I_k,\J_k}, \epsilon_{\I_l,\J_l}),
\end{equation}
and $X \hadprod Y$ is the Hadamard (or element-wise) product
\begin{equation}
  (X \hadprod Y)_{i,j} = X_{i,j} Y_{i,j}.
\end{equation}
We now assume that the GP $G$ can be modeled by a semiseparable kernel (see Eq.~(\ref{eq:semisep})).
We can thus use the merged time vector $t$ to compute the semiseparable representation of
the covariance matrix of $G(t)$, $G'(t)$ according to
Eqs.~(\ref{eq:semisep}),~(\ref{eq:celeritedecomp}),~(\ref{eq:DD2}), and (\ref{eq:Dceleritedecomp})
(see also Appendix~\ref{sec:matern}),
and we obtain
\begin{align}
  \label{eq:multisemisep_1}
  C =
   & \ \l(\vec{\alpha}\vec{\alpha}\t\r) \hadprod
  \l( \diag(A) + \tril\l(U V\t\r) + \triu\l(V U\t\r) \r)\nonumber         \\
   & + \l(\vec{\beta} \vec{\alpha}\t\r) \hadprod
  \l( \tril\l(U' V\t\r) + \triu\l(V' U\t\r) \r)\nonumber                  \\
   & + \l(\vec{\alpha} \vec{\beta}\t\r) \hadprod
  \l( \tril\l(U {V'}\t\r) + \triu\l(V {U'}\t\r) \r)\nonumber              \\
   & + \l(\vec{\beta} \vec{\beta}\t\r) \hadprod
  \l( \diag(B) + \tril\l(U' {V'}\t\r) + \triu\l(V' {U'}\t\r) \r)\nonumber \\
   & + \Sigma.
\end{align}
The Hadamard product $\l(\vec\alpha\vec\beta\t\r) \hadprod M$ can also be rewritten as
\begin{equation}
  \l(\vec\alpha\vec\beta\t\r) \hadprod M = \diag(\vec\alpha) M \diag(\vec\beta),
\end{equation}
thus
\begin{align}
   & \l(\vec{\alpha}\vec{\alpha}\t\r) \hadprod \diag(A) = \diag\l(\vec{\alpha}^2*A\r),\nonumber                                \\
   & \l(\vec{\alpha}\vec{\beta}\t\r) \hadprod \tril\l(UV\t\r) = \tril\l(\l(\vec{\alpha}*U\r)\l(\vec{\beta}*V\r)\t\r),\nonumber \\
   & \l(\vec{\alpha}\vec{\beta}\t\r) \hadprod \triu\l(UV\t\r) = \triu\l(\l(\vec{\alpha}*U\r)\l(\vec{\beta}*V\r)\t\r),
\end{align}
where $\vec\alpha^2$ is the element-wise square of $\vec\alpha$ ($\vec\alpha^2 = \vec\alpha*\vec\alpha$)
and $\vec\alpha * U$ is the element-wise product of each column of $U$ by $\vec\alpha$
\begin{equation}
  \l(\vec\alpha * U\r)_{k,s} = \alpha_k U_{k,s}.
\end{equation}
Using these relations, Eq.~(\ref{eq:multisemisep_1}) can be rewritten as
\begin{align}
  \label{eq:multisemisepexpand}
  C = & \ \diag\l(\vec{\alpha}^2 \hadprod A\r)\nonumber                            \\
      & + \tril\l(\l(\vec{\alpha}\hadprod U\r)\l(\vec{\alpha}\hadprod V\r)\t\r)
  + \triu\l(\l(\vec{\alpha}\hadprod V\r)\l(\vec{\alpha}\hadprod U\r)\t\r)\nonumber \\
      & + \tril\l(\l(\vec{\beta}\hadprod U'\r)\l(\vec{\alpha}\hadprod V\r)\t\r)
  + \triu\l(\l(\vec{\beta}\hadprod V'\r)\l(\vec{\alpha}\hadprod U\r)\t\r)\nonumber \\
      & + \tril\l(\l(\vec{\alpha}\hadprod U\r)\l(\vec{\beta}\hadprod V'\r)\t\r)
  + \triu\l(\l(\vec{\alpha}\hadprod V\r)\l(\vec{\beta}\hadprod U'\r)\t\r)\nonumber \\
      & + \diag\l(\vec{\beta}^2 \hadprod B\r)\nonumber                             \\
      & + \tril\l(\l(\vec{\beta}\hadprod U'\r)\l(\vec{\beta}\hadprod V'\r)\t\r)
  + \triu\l(\l(\vec{\beta}\hadprod V'\r)\l(\vec{\beta}\hadprod U'\r)\t\r)\nonumber \\
      & + \Sigma.
\end{align}
Finally, the latter expression can be factorized to obtain the
semiseparable representation
\begin{equation}
  \label{eq:multisemisep}
  C = \diag(\A) + \tril\l(\U \V\t\r) + \triu\l(\V \U\t\r) + \Sigma,
\end{equation}
with
\begin{align}
  \label{eq:celerite2decomp}
  \A & = \vec{\alpha}^2 \hadprod A + \vec{\beta}^2 \hadprod B,\nonumber \\
  \U & = \vec{\alpha}\hadprod U + \vec{\beta}\hadprod U',\nonumber      \\
  \V & = \vec{\alpha}\hadprod V + \vec{\beta}\hadprod V'.
\end{align}
The rank of this semiseparable representation of $C$ (number of columns in $\U$, $\V$)
is thus the same as the rank of the underlying GP (number of columns in $U$, $V$).
In the case of the more general model of Eq.~(\ref{eq:RajpaulGeneral}),
with several independent GP,
it is straightforward to compute the semiseparable representation of the covariance matrix
by vertical concatenation of the matrix $\U$ and $\V$
corresponding to each independent GP
and the total rank is the sum of all the GP ranks.

Keeping the rank of the covariance matrix as low as possible allows
us to significantly improve the performances of the method.
Indeed, the cost of likelihood evaluations with a
semiseparable covariance matrix of rank $r$ scales as $\O\l(n r^2\r)$.
It is thus remarkable to note that introducing the derivative of the GP in the model,
as well as different coefficients $\alpha$ and $\beta$ for the different time series,
does not increase the rank of the semiseparable representation of the covariance matrix.
The factorization performed between Eq.~(\ref{eq:multisemisepexpand})
and Eqs.~(\ref{eq:multisemisep}) and~(\ref{eq:celerite2decomp})
is the key step that allows us to keep the same rank as the underlying GP.
This factorization is achieved thanks to the specific choice of semiseparable representation
introduced in Eq.~(\ref{eq:DD2}).
Indeed, instead of using $(U, V, U', V')$ to represent the covariance matrix,
one could use $(U, V, U', U'')$ or $(U, V, V', V'')$
(see Eqs.~(\ref{eq:Dantisep}),~(\ref{eq:D2semisep})).
However, with these choices of representation,
the covariance matrix would not factorize in the same way,
the rank of its semiseparable representation would be twice the rank of the underlying GP,
and the cost would thus quadruple.

\subsection{Measurement noise and \leaf{} component}
\label{sec:spleaf2}

Using the covariance matrix representation of Eq.~(\ref{eq:multisemisep}),
the results described in Sect.~\ref{sec:homogeneous} in the case of a homogeneous time series
can be extended to the case of heterogeneous time series depending on several independent GP
and their derivative.
The \celerite{} algorithms can be applied in the case of purely white noise ($\Sigma$ diagonal)
and the \spleaf{} algorithms can be applied to the more general case of close-to-diagonal correlated noise.
The \leaf{} component of the \spleaf{} model simply needs to be defined on the merged time series $(t,y)$.

\subsection{Computational cost and memory footprint}
\label{sec:cost}

The computational cost of our model and its memory footprint
are the same as the underlying \celerite{}/\spleaf{} representation.
Therefore, the memory footprint scales as $\O\l(\l(r+\bar{b}\r)n\r)$
and its computational cost as $\O\l(\l(r^2+r\bar{b}+\bar{b}^2\r)n\r)$,
where
$n$ is the total number of measurements (including radial velocities and indicators),
$r$ is the rank of $\U$ and $\V$,
$\bar{b}$ is the average band width of the $\leaf{}$ component (considering the merged time series $(t,y)$).
This is to be compared with a naive implementation of the same model
which has a memory footprint in $\O\l(n^2\r)$ and a computational cost in $\O\l(n^3\r)$.

\subsection{Overflows and preconditioning}
\label{sec:overflows}

As explained in \citet{ambikasaran_2015_generalized,foreman-mackey_2017_fast,delisle_2020_efficientb},
a naive computer implementation of the semiseparable decomposition
of Eq.~(\ref{eq:celeritedecomp})
can lead to overflows and underflows due to the exponential terms in the definition of $U$ and $V$.
However, \citet{foreman-mackey_2017_fast} proposed a simple preconditioning method to circumvent this issue
in the case of the \celerite{} model,
which is also valid in the case of the \spleaf{} model \citep{delisle_2020_efficientb}.
Instead of using directly the matrices $U$ and $V$, one might use the matrices
$\tilde{U}$, $\tilde{V}$, and $\phi$ defined as
\begin{align}
  \tilde{U}_{i,s}              & = a_s \cos(\nu_s t_i) + b_s \sin(\nu_s t_i),\nonumber       \\
  \tilde{U}_{i,n_\mathrm{c}+s} & = a_s \sin(\nu_s t_i) - b_s\cos(\nu_s t_i),\nonumber        \\
  \tilde{V}_{i,s}              & = \cos(\nu_s t_i),\nonumber                                 \\
  \tilde{V}_{i,n_\mathrm{c}+s} & = \sin(\nu_s t_i),\nonumber                                 \\
  \phi_{i,s}                   & = \phi_{i,n_\mathrm{c}+s} = \expo{-\lambda_s(t_{i+1}-t_i)},
\end{align}
such that
\begin{equation}
  U_{i,s} V_{j,s} = \tilde{U}_{i,s} \tilde{V}_{j,s} \prod_{k=j}^{i-1} \phi_{k,s},
\end{equation}
and all the \celerite{}/\spleaf{} algorithms can be adapted to use this representation
\citep[see][]{foreman-mackey_2017_fast,foreman-mackey_2018_scalable,delisle_2020_efficientb}.
Similarly, we define the matrices $\tilde{U}'$, $\tilde{V}'$, $\tilde{\U}$, and $\tilde{\V}$ as
\begin{align}
  \tilde{U}'_{i,s}              & = a'_s \cos(\nu_s t_i) + b'_s \sin(\nu_s t_i),\nonumber                     \\
  \tilde{U}'_{i,n_\mathrm{c}+s} & = a'_s \sin(\nu_s t_i) - b'_s\cos(\nu_s t_i),\nonumber                      \\
  \tilde{V}'_{i,s}              & = \lambda_s \cos(\nu_s t_i) - \nu_s \sin(\nu_s t_i),\nonumber               \\
  \tilde{V}'_{i,n_\mathrm{c}+s} & = \lambda_s \sin(\nu_s t_i) + \nu_s\cos(\nu_s t_i),\nonumber                \\
  \tilde{\U}                    & = \vec{\alpha}\hadprod \tilde{U} + \vec{\beta}\hadprod \tilde{U}',\nonumber \\
  \tilde{\V}                    & = \vec{\alpha}\hadprod \tilde{V} + \vec{\beta}\hadprod \tilde{V}',
\end{align}
which allow to apply the overflow-proof version of \celerite{}/\spleaf{} algorithms.
The same preconditioning method can be applied in the case of the Matérn 3/2 and 5/2 kernels
(see Appendix~\ref{sec:matern_overflows}).

\subsection{Efficient computation of the gradient}
\label{sec:gradient}

In most applications, one needs to explore the
parameter space either to maximize the likelihood using optimization algorithms
or to obtain samples from the posterior distribution of parameters using Bayesian methods (MCMC, nested sampling, etc.).
In both cases, many algorithms have been designed that make use of the gradient of the likelihood
with respect to the parameters to improve the convergence efficiency.
Following \citet{foreman-mackey_2018_scalable} and \citet{delisle_2020_efficientb},
we deduce gradient backpropagation algorithms for all the operations used to compute the likelihood.
While a detailed presentation of these backpropagation algorithms would be cumbersome,
we refer the reader to \citet{delisle_2020_efficientb} Appendix~B for the general idea of the method,
and to the reference \spleaftwo{} implementation\footnote{\spleafURL} for further details.

\section{Application: Reanalysis of HD~13808}
\label{sec:application}

In this section, we apply our algorithms to reanalyze
the RV time series of \object{HD~13808}.
This K2V dwarf is known to harbor two planet candidates \citep[see][]{mayor_2011_harps}
recently published as confirmed planets by \citet{ahrer_2021_harps}.
In the latter study, the authors defined several alternative models of stellar activity
and performed a Bayesian model comparison letting the number of planets vary.
In addition to confirming the two candidates,
\citet{ahrer_2021_harps} concluded that the best stellar activity model
was a GP trained simultaneously on the RV, BIS, and $\rhk$ time series,
following Eq.~(\ref{eq:RajpaulModel}), as proposed by \citet{rajpaul_2015_gaussian}.
Modeling this GP required the authors to solve for a $738 \times 738$ linear system billions of times,
which was very demanding in terms of computational resources.
Here, we reanalyze the 246 HARPS RV measurements of HD~13808,
together with the indicators time series,
but modeling the GP using \spleaftwo{}.

\subsection{Choice of kernel}
\label{sec:kernel}

The SEP kernel (see Eq.~(\ref{eq:SEPkernel})),
which was used by \citet{ahrer_2021_harps} for their analysis of HD~13808,
is not semiseparable and thus cannot be modeled with \spleaftwo{}.
However, other quasiperiodic kernels, such as the SHO kernel proposed by \citet{foreman-mackey_2017_fast},
admit a semiseparable representation.
Here, we aim to reproduce the main characteristics of the SEP kernel but using a semiseparable kernel.
The SEP kernel is the product of a squared-exponential and the exponential of a sinusoidal
(see Eq.~(\ref{eq:SEPkernel})):
\begin{equation}
  k(\Delta t) = \sigma^2 \exp \l(- \frac{\Delta t^2}{2 \rho^2}\r)
  \exp\l(-\frac{\sin^2 \l( \frac{\pi \Delta t }{P}\r) }{2 \eta^2}\r).
\end{equation}
The second part can be expanded as a power series, assuming $2 \eta \gtrsim 1$
\begin{align}
  \label{eq:SEPdev}
  k(\Delta t) = & \ \sigma^2 \exp \l(- \frac{\Delta t^2}{2 \rho^2}\r)
  \l(  1 - \frac{\sin^2 \l( \frac{\pi \Delta t }{P}\r) }{2 \eta^2}
  + \frac{\sin^4 \l( \frac{\pi \Delta t }{P}\r) }{8\eta^4} + \O\l(\eta^{-6}\r)\r)\nonumber \\
  =             & \ \sigma^2 \exp \l(- \frac{\Delta t^2}{2 \rho^2}\r)
  \frac{1 + f\cos\l(\nu \Delta t\r)
    + \frac{f^2}{4} \cos\l(2\nu\Delta t\r)}{1+f+\frac{f^2}{4}}
  + \O\l(f^3\r),
\end{align}
with $f=(2\eta)^{-2}$ and $\nu = 2\pi/P$.
This kernel thus introduces some correlation at the rotation period $P$,
but also at the harmonics ($P/2$, $P/3$, etc.),
with amplitudes decaying rapidly (scaling as $\eta^{-2n}$ for the harmonics $P/n$).
The squared-exponential part implies that the correlations vanish over long timescales ($\Delta t \gg \rho$).
The squared-exponential kernel is not semiseparable,
but the Matérn 1/2 (simple exponential decay), 3/2, and 5/2 kernels
admit a semiseparable decomposition (see Appendix~\ref{sec:matern}),
and offer a similar decay of the correlation over long timescales.
The SEP kernel could thus be roughly approximated by:
\begin{equation}
  k(\Delta t) = \sigma^2 \exp\l(-\frac{\Delta t}{\rho}\r)
  \frac{1 + f\cos\l(\nu \Delta t\r)
    + \frac{f^2}{4} \cos\l(2\nu\Delta t\r)}{1+f+\frac{f^2}{4}}.
\end{equation}
However, a GP following this kernel would not be differentiable ($k'(0)=-\sigma^2/\rho \neq 0$).
In order to ensure differentiability, we introduce a modified kernel,
which is the combination of a Matérn 3/2 kernel and two underdamped SHO terms
\begin{equation}
  \label{eq:MEPkernel}
  k(\Delta t) = \sigma^2\frac{k_{3/2}(\Delta t)
    + f k_{\mathrm{SHO,\,fund.}}(\Delta t)
    + \frac{f^2}{4} k_{\mathrm{SHO,\,harm.}}(\Delta t)}{1+f+\frac{f^2}{4}},
\end{equation}
where
\begin{align}
  k_{3/2}(\Delta t)                  & = \exp\l(-\frac{\sqrt{3}\Delta t}{\rho}\r)\l(1+\frac{\sqrt{3}\Delta t}{\rho}\r),\nonumber                        \\
  k_{\mathrm{SHO,\,fund.}}(\Delta t) & = \exp\l(-\frac{\Delta t}{\rho}\r)\l(\cos\l(\nu \Delta t\r)+\frac{1}{\nu\rho}\sin\l(\nu \Delta t\r)\r),\nonumber \\
  k_{\mathrm{SHO,\,harm.}}(\Delta t) & = \exp\l(-\frac{\Delta t}{\rho}\r)\l(\cos\l(2\nu \Delta t\r)+\frac{1}{2\nu\rho}\sin\l(2\nu \Delta t\r)\r).
\end{align}
The kernel of Eq.~(\ref{eq:MEPkernel}),
which we refer to as the Matérn 3/2 exponential periodic (MEP) kernel
in the following,
is differentiable and presents the
main characteristics of the SEP kernel,
while being semiseparable.
The semiseparable representation of the MEP kernel is of rank $r=6$.
We note that similar kernels have already been used in the literature
to model stellar activity in photometric time series
\citep[e.g.,][]{david_2019_four,gillen_2020_ngts}.

It should be noted that the MEP kernel is once mean square differentiable but not twice,
which means that $G'$ is well defined but it is not itself differentiable.
Since the time series (RV and activity indicators) are modeled as combinations
of $G$ and $G'$,
we could require $G'$ to be differentiable to obtain a smoother model.
Such a twice mean square differentiable kernel should satisfy $k^{(3)}(0) = 0$,
in addition to the mandatory differentiability condition ($k'(0) = 0$).
For the MEP kernel, $k_\mathrm{MEP}^{(3)}(0)$ is non-zero,
however, in practice this kernel seems to produce smooth time series
(see Fig.~\ref{fig:residuals_zoom}).
Nevertheless, it is possible to design semiseparable kernels
that are rigorously twice differentiable.
For instance, the Matérn 5/2 kernel is twice differentiable
and semiseparable with a rank of 3 (see Appendix~\ref{sec:matern}).
In Appendix~\ref{sec:twicediff} we present the exponential-sine (ES)
and the exponential-sine periodic (ESP) kernels.
Both kernels are twice differentiable and semiseparable.
The ES kernel is of rank 3 and closely resemble the SE kernel,
while the ESP kernel is of rank 15 and approximates the SEP kernel very well.
We find very similar results when using the ESP kernel instead of the MEP kernel,
while the cost of likelihood evaluations is roughly doubled
because of the higher rank of the ESP kernel
(see Appendix~\ref{sec:twicediff}).
In the following, we thus adopt the MEP kernel as a replacement for the SEP kernel,
to reduce the computational cost and since it produces smooth time series,
at least in our case.
In the general case, the ESP kernel would typically
generate smoother times series than the MEP,
but with a doubled cost.

Following \citet{rajpaul_2015_gaussian} and \citet{ahrer_2021_harps} we
use the GP to model
simultaneously the RV, BIS, and $\rhk$ time series of HD~13808
according to (see Eq.~(\ref{eq:RajpaulModel})):
\begin{align}
  \label{eq:GPmodel}
   & \Delta \mathrm{RV} = \alpha_\mathrm{RV} G(t) + \beta_\mathrm{RV} G'(t),\nonumber    \\
   & \Delta \mathrm{BIS} = \alpha_\mathrm{BIS} G(t) + \beta_\mathrm{BIS} G'(t),\nonumber \\
   & \Delta \rhk = \alpha_{\rhk} G(t),
\end{align}
where the coefficients $\alpha$, $\beta$ and the kernel's hyperparameters ($P$, $\rho$, $\eta$)
need to be determined.

\subsection{Performances}
\label{sec:perfs}

\begin{figure}
  \centering
  \includegraphics[width=\linewidth]{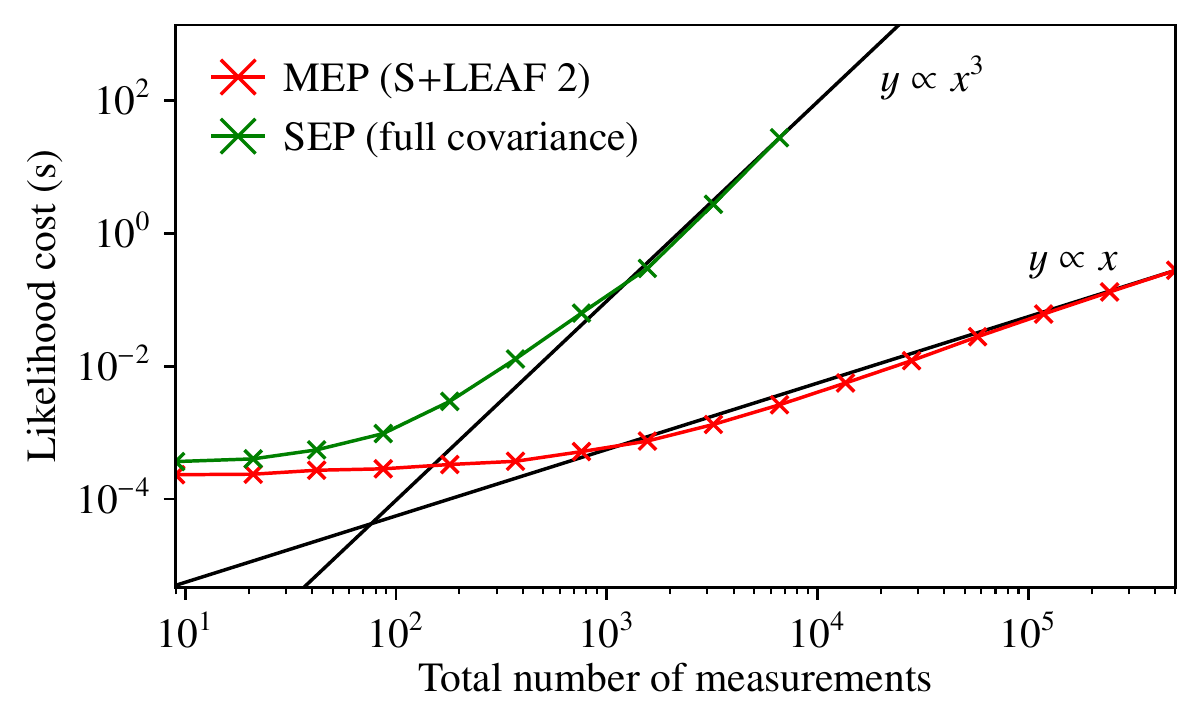}
  \caption{Cost of a likelihood evaluation as a function of the total number of measurements
    using \spleaftwo{} or the full covariance matrix (see Sect.~\ref{sec:perfs}).
  }
  \label{fig:perfs}
\end{figure}

\begin{figure}
  \centering
  \includegraphics[width=\linewidth]{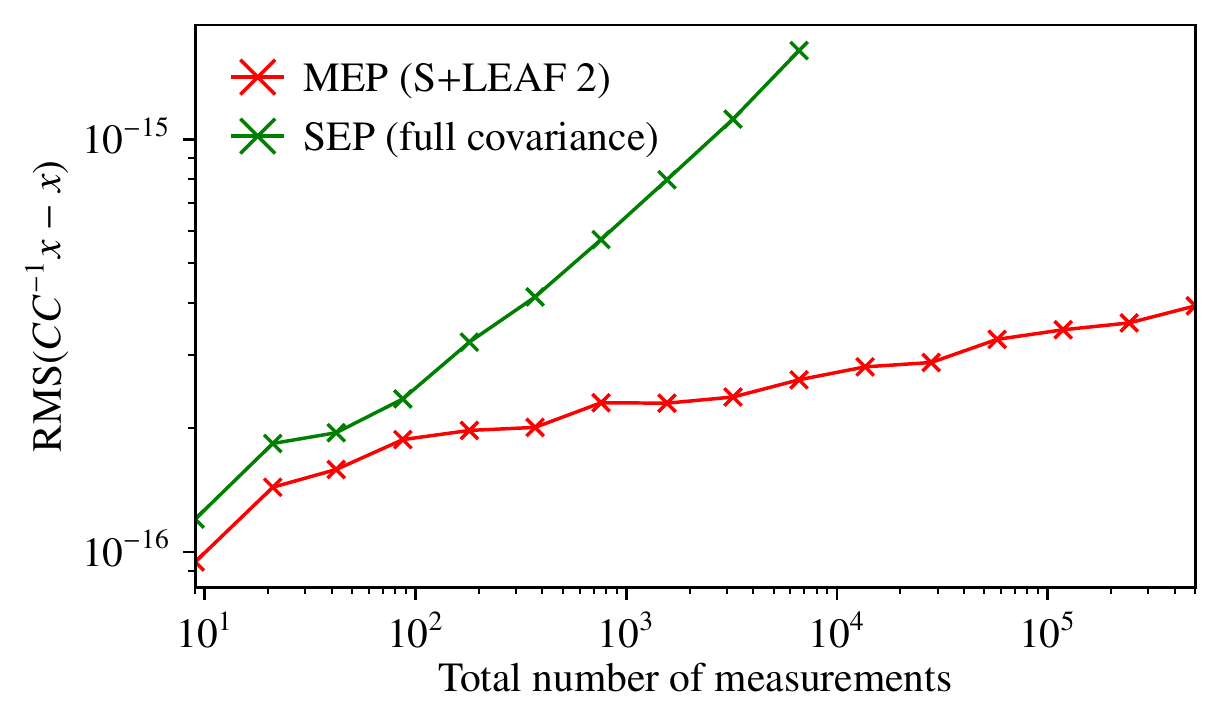}
  \caption{Precision of the linear solving operation as a function of the total number of measurements
    using \spleaftwo{} or the full covariance matrix (see Sect.~\ref{sec:perfs}).
  }
  \label{fig:prec}
\end{figure}

To evaluate the performances of \spleaftwo{} for the model of Eqs.~(\ref{eq:MEPkernel}),~(\ref{eq:GPmodel}),
we generated random RV, BIS, and $\rhk$ time series
and record the cost of the likelihood evaluation
as a function of the number of generated measurements.
We compared \spleaftwo{} performances using the MEP kernel
with the computational cost of evaluating the likelihood using the full covariance matrix
with the SEP kernel.
The two implementations were run on the same computer, using a single core.
The results are shown in Fig.~\ref{fig:perfs}
and confirm the $\O\l(n\r)$ scaling of \spleaftwo{},
while the implementation of the full covariance matrix indeed scales as $\O\l(n^3\r)$ for a large value of $n$.
In the case of HD~13808, the total number of measurements is 738 ($3\times 246$),
and based on Fig.~\ref{fig:perfs},
we obtain a gain of a factor $\sim 130$ in computing time
by using \spleaftwo{} instead of the naive implementation.
For larger data sets, the gain would be even greater.

In addition to these performances tests, we also ran
numerical precision tests.
We used the same GP model as for the performances tests
and assessed the stability of \spleaftwo{} by
computing $C C^{-1} x$, that is, applying the solving and dot product algorithms
on a random merged time series $x$.
In theory, we should find $C C^{-1} x = x$,
however, due to the limited machine precision,
numerical errors accumulate and the results stray slightly from $x$.
We computed the root mean square (rms) of these errors,
for the \spleaftwo{} methods and for the full covariance matrix
(see Fig.~\ref{fig:prec}).
In both cases, numerical errors grow with the total number of measurements.
However, the level and growth rate are lower when using \spleaftwo{}.
For a given number of measurements,
the number of arithmetic operations is lower when using \spleaftwo{},
which could explain the improvement in precision.

\subsection{Periodogram and false alarm probability}
\label{sec:periofap}

\begin{figure}
  \centering
  \includegraphics[width=\linewidth]{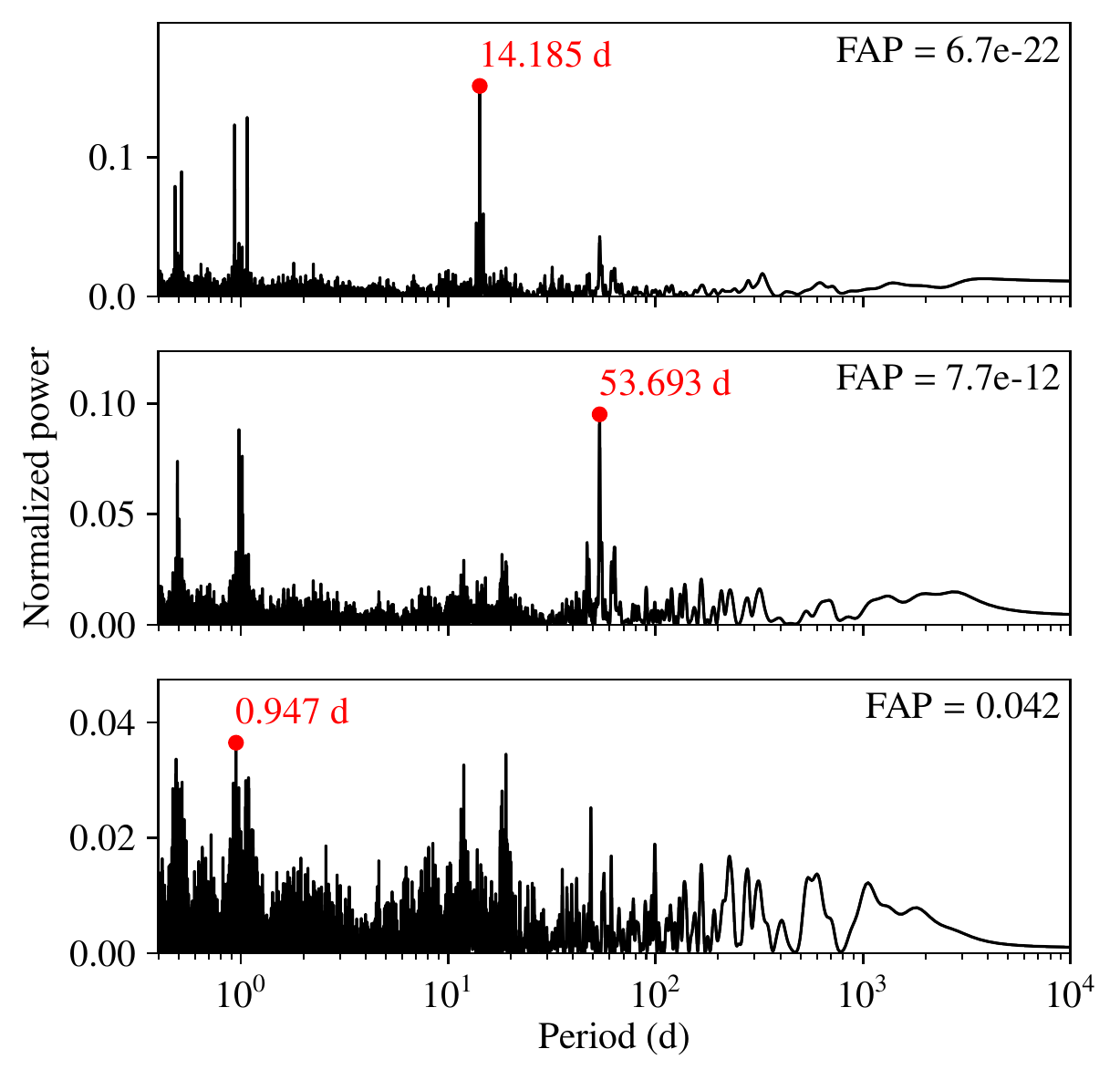}
  \caption{Periodograms of the raw RV time series of HD~13808 (\textit{top})
    as well as of the residuals after subtracting the 14.19~d (\textit{center})
    and the 53.7~d (\textit{bottom}) planets.
  }
  \label{fig:perio}
\end{figure}

\begin{figure}
  \centering
  \includegraphics[width=\linewidth]{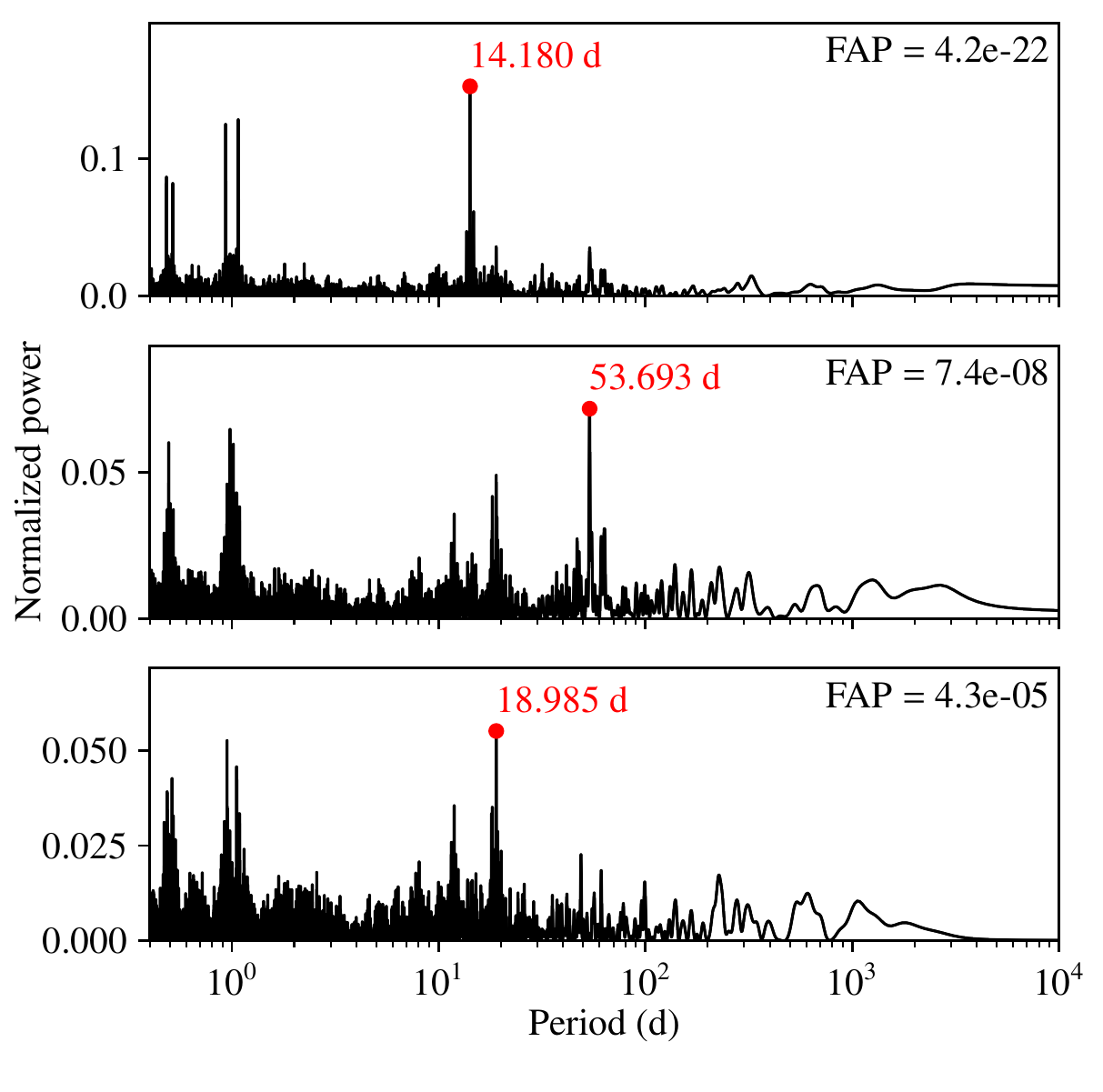}
  \caption{Same as Fig.~\ref{fig:perio} but neglecting the harmonics component
    of the MEP kernel (see Eq.~(\ref{eq:MEPkernel})).
  }
  \label{fig:perio_noharm}
\end{figure}

\begin{table}
  \caption{Maximum likelihood solution and \textsc{PolyChord} posterior for the model with a GP and two planets (at 14.19 and 53.7~d).}
  \centering
  \renewcommand{\arraystretch}{1.4}\setlength{\tabcolsep}{2.5pt}
  \begin{tabular}{cccc}
    \hline
    \hline
    Parameter                      & (units)                        & Maximum likelihood      & \textsc{PolyChord} posterior       \\ \hline
    $P_\mathrm{GP}$                & (d)                            & $38.171 \pm 0.246$      & $38.162^{+0.159}_{-0.149}$         \\
    $\rho_\mathrm{GP}$             & (d)                            & $279.6 \pm 59.0$        & $308.64^{+54.5}_{-57.5}$           \\
    $\eta_\mathrm{GP}$             &                                & $0.978 \pm 0.125$       & $1.053^{+0.146}_{-0.131}$          \\ \hline
    $\gamma_\mathrm{RV}$           & ($\mathrm{m} \mathrm{s}^{-1}$) & $41095.100 \pm 0.419$   & $41095.132^{+0.507}_{-0.516}$      \\
    $\sigma_\mathrm{RV}$           & ($\mathrm{m} \mathrm{s}^{-1}$) & $1.700 \pm 0.113$       & $1.815^{+0.123}_{-0.117}$          \\
    $\alpha_\mathrm{RV}$           & ($\mathrm{m} \mathrm{s}^{-1}$) & $1.157 \pm 0.198$       & $1.373^{+0.274}_{-0.221}$          \\
    $\beta_\mathrm{RV}$            &                                & $20.55 \pm 4.64$        & $22.123^{+5.64}_{-5.00}$           \\ \hline
    $\gamma_\mathrm{BIS}$          & ($\mathrm{m} \mathrm{s}^{-1}$) & $5.605 \pm 0.939$       & $5.583^{+1.12}_{-1.11}$            \\
    $\sigma_\mathrm{BIS}$          & ($\mathrm{m} \mathrm{s}^{-1}$) & $1.830 \pm 0.140$       & $1.775^{+0.159}_{-0.149}$          \\
    $\alpha_\mathrm{BIS}$          & ($\mathrm{m} \mathrm{s}^{-1}$) & $2.699 \pm 0.408$       & $3.104^{+0.484}_{-0.428}$          \\
    $\beta_\mathrm{BIS}$           &                                & $-27.08 \pm 5.31$       & $-31.597^{+5.80}_{-6.70}$          \\ \hline
    $\gamma_{\log R_{HK}'}$        &                                & $-4.8774 \pm 0.0181$    & $-4.8780^{+0.0215}_{-0.0218}$      \\
    $\sigma_\mathrm{\log R_{HK}'}$ &                                & $0.007374 \pm 0.000759$ & $0.007751^{+0.000856}_{-0.000804}$ \\
    $\alpha_{\log R_{HK}'}$        &                                & $0.05273 \pm 0.00767$   & $0.06050^{+0.00907}_{-0.00803}$    \\ \hline
    $P_\mathrm{b}$                 & (d)                            & $14.18538 \pm 0.00199$  & $14.18657^{+0.00218}_{-0.00224}$   \\
    $\lambda_\mathrm{0,\,b}$       & (deg)                          & $-49.55 \pm 3.30$       & $-47.59^{+3.65}_{-3.71}$           \\
    $K_\mathrm{b}$                 & ($\mathrm{m} \mathrm{s}^{-1}$) & $3.669 \pm 0.194$       & $3.677^{+0.207}_{-0.206}$          \\
    $e_\mathrm{b}$                 &                                & $0.0759 \pm 0.0544$     & $0.0441^{+0.0477}_{-0.0304}$       \\
    $\omega_\mathrm{b}$            & (deg)                          & $213.6 \pm 34.4$        & $224.1^{+73.2}_{-107}$             \\ \hline
    $P_\mathrm{c}$                 & (d)                            & $53.6956 \pm 0.0437$    & $53.7126^{+0.0533}_{-0.0514}$      \\
    $\lambda_\mathrm{0,\,c}$       & (deg)                          & $-8.64 \pm 6.04$        & $-6.561^{+6.54}_{-6.65}$           \\
    $K_\mathrm{c}$                 & ($\mathrm{m} \mathrm{s}^{-1}$) & $2.027 \pm 0.191$       & $1.948^{+0.199}_{-0.202}$          \\
    $e_\mathrm{c}$                 &                                & $0.165 \pm 0.117$       & $0.0929^{+0.101}_{-0.0648}$        \\
    $\omega_\mathrm{c}$            & (deg)                          & $280.7 \pm 32.0$        & $251.6^{+60.2}_{-137}$             \\
    \hline
  \end{tabular}
  \tablefoot{The reference epoch is 2\,455\,000~BJD.
    The 1-$\sigma$ uncertainties provided together with the maximum likelihood solution
    are derived from the inverse of the Hessian matrix of the log-likelihood.
    The \textsc{PolyChord} posterior values and uncertainties correspond to the median and
    the 15.865 and 84.135 percentiles.}
  \label{tab:params}
\end{table}

\begin{figure}
  \centering
  \includegraphics[width=\linewidth]{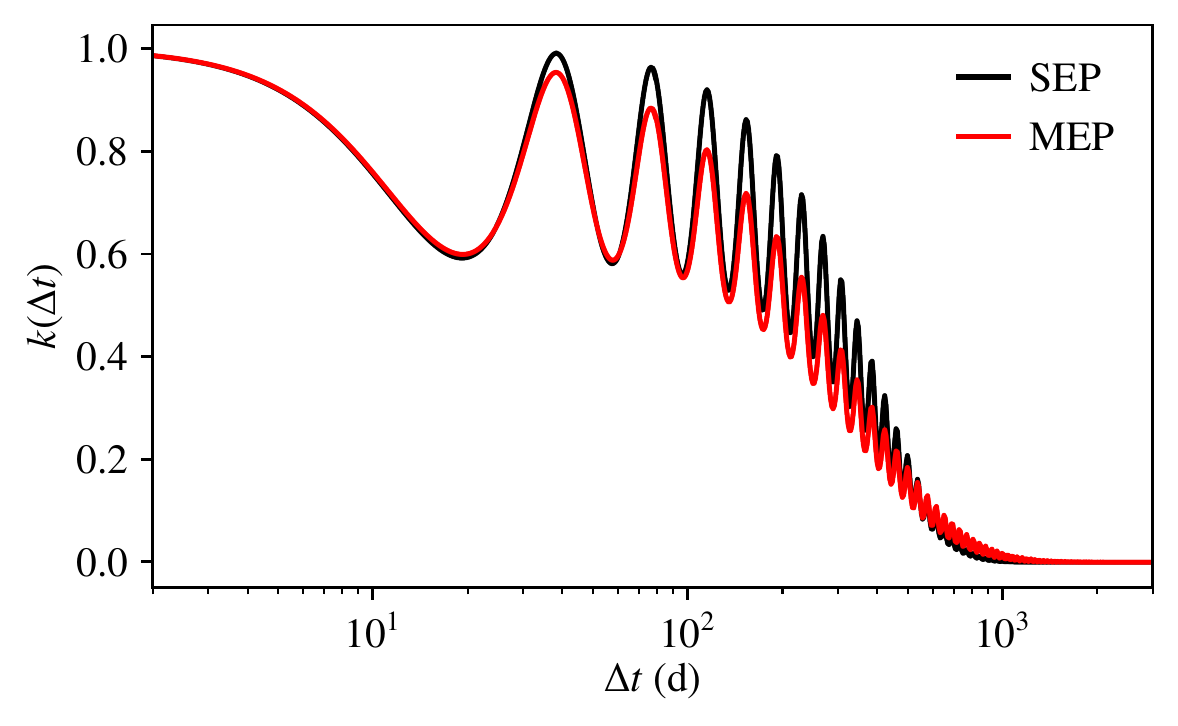}
  \caption{Kernel function used to model HD~13808's activity (MEP, see Eq.~(\ref{eq:MEPkernel})).
    The GP hyperparameters are taken from the best fit of the two planets solution (Table~\ref{tab:params}).
    For comparison, the SEP kernel, which the MEP kernel is design to roughly mimic,
    is also plotted using the same set of hyperparameters.}
  \label{fig:GP_kernel}
\end{figure}

\begin{figure}
  \centering
  \includegraphics[width=\linewidth]{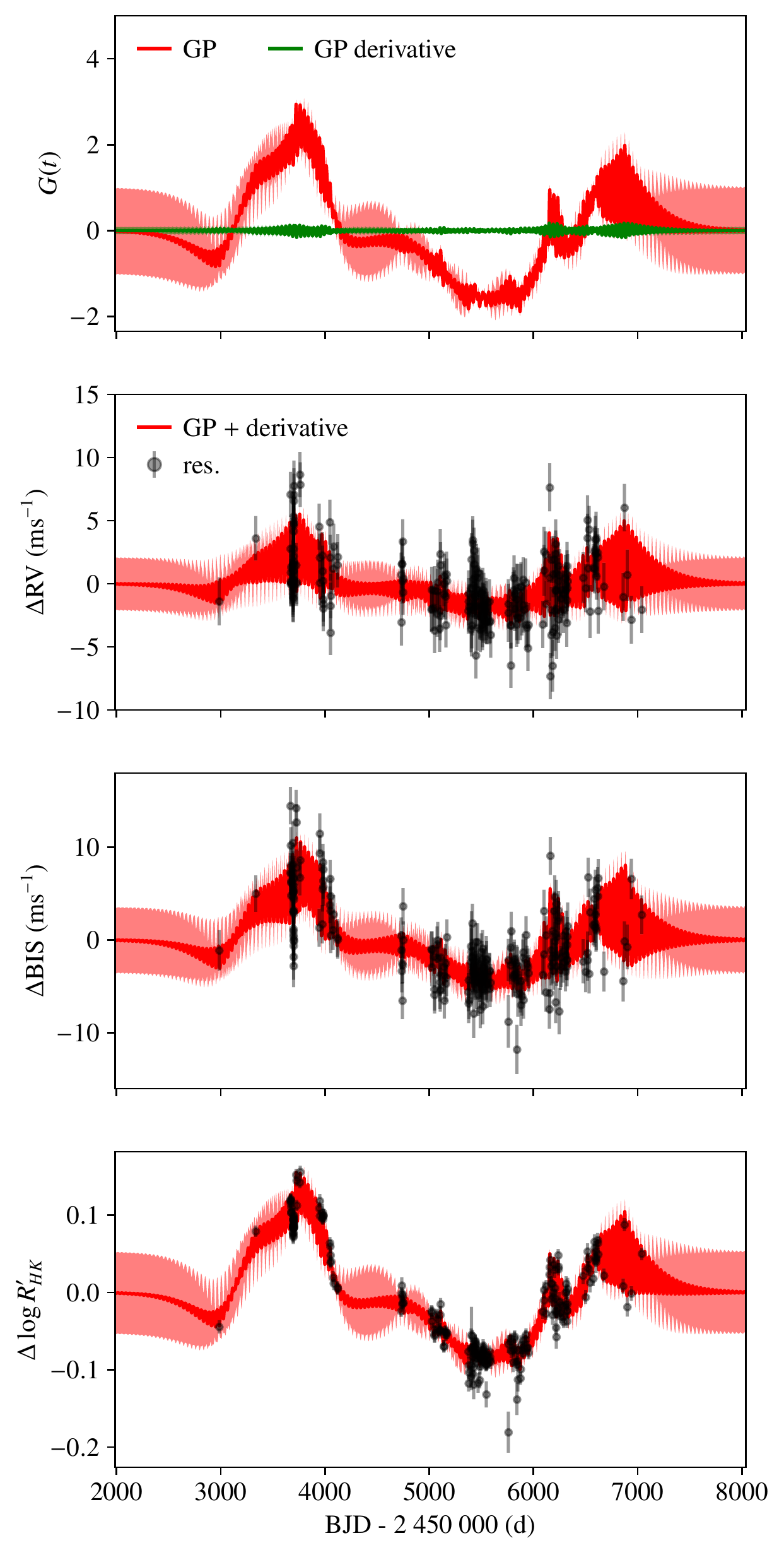}
  \caption{GP prediction (conditional distribution) from
    the best fit of the two planets solution (Table~\ref{tab:params}).
    The prediction is plotted for the GP and its derivative (\textit{top})
    and the full GP prediction for the RV, BIS, and $\rhk$ time series
    superimposed with the corresponding residuals
    (\textit{bottom three plots}).
  }
  \label{fig:residuals}
\end{figure}

\begin{figure}
  \centering
  \includegraphics[width=\linewidth]{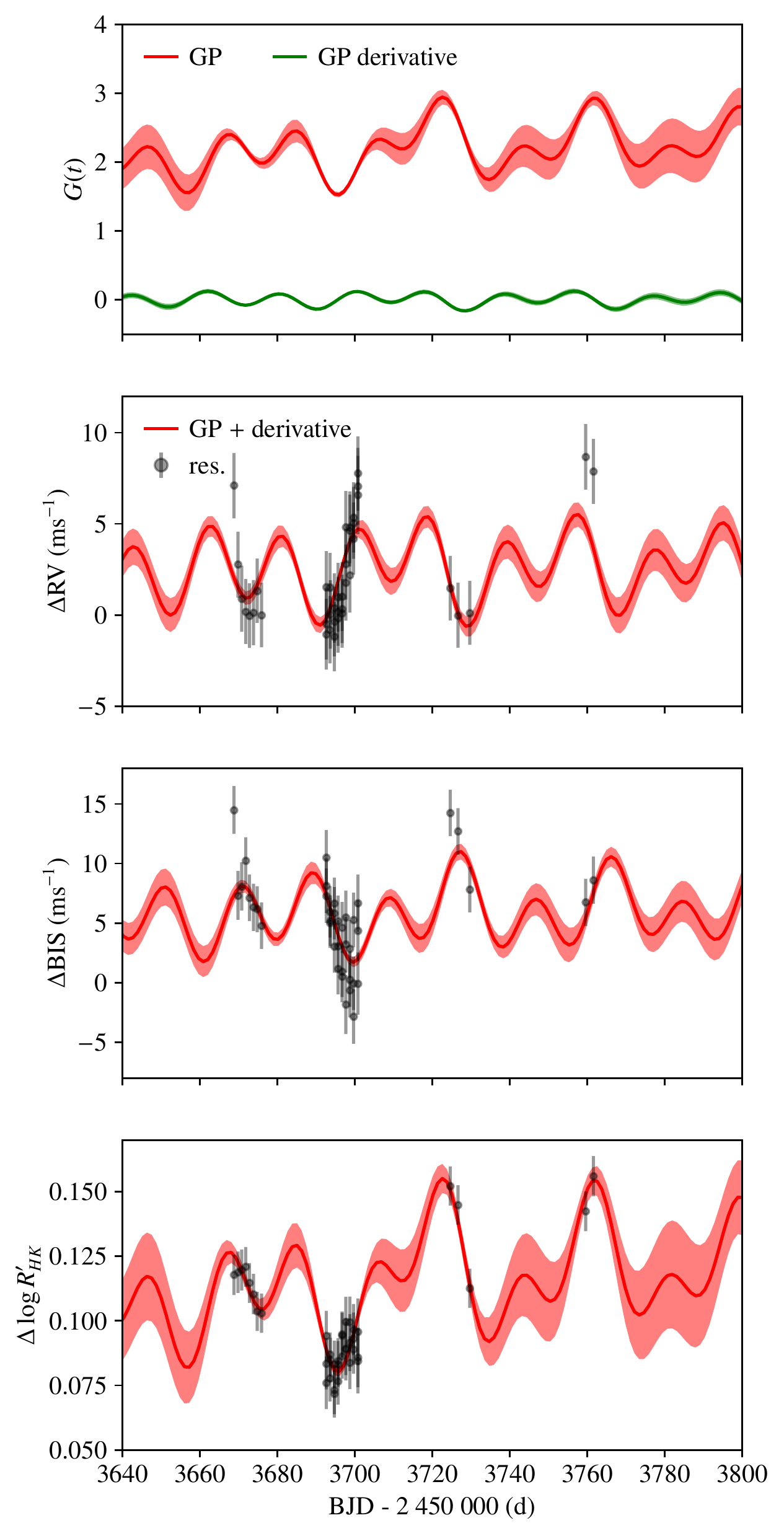}
  \caption{Zoom of Fig.~\ref{fig:residuals} around epoch 2\,453\,720~BJD.}
  \label{fig:residuals_zoom}
\end{figure}

We analyzed HD~13808 data using the periodogram and false alarm probability (FAP) approach
presented in \citet{delisle_2020_efficient}.
For each frequency, all the linear parameters (here the RV and indicators offsets)
are re-adjusted together with the amplitudes of the cosine and sine at the considered frequency
(only applied to the RV time series).
The framework of \citet{delisle_2020_efficient} only requires slight adaptations to account
for the joint fit of RV and activity indicators, which are detailed in Appendix~\ref{sec:fapmodif}.

We first performed a fit of a base model including the offsets
$\gamma_\mathrm{RV}$, $\gamma_\mathrm{BIS}$, $\gamma_{\rhk}$,
jitter terms added in quadrature to the measurements errorbars,
$\sigma_\mathrm{RV}$, $\sigma_\mathrm{BIS}$, $\sigma_{\rhk}$,
the kernels hyperparameters,
$P_\mathrm{GP}$, $\rho_\mathrm{GP}$, $\eta_\mathrm{GP}$,
and the amplitudes, $\alpha$, $\beta$,
by maximizing the likelihood.
Then using the fitted noise parameters, we computed a first periodogram,
reajusting the offsets for each considered frequency.
The resulting periodogram is plotted in Fig.~\ref{fig:perio} (top).
We observe a very significant peak at 14.19~d ($\mathrm{FAP} = 6.7\times 10^{-22}$).
We then fitted a Keplerian orbit to this signal, and reajusted all parameters.
A second periodogram (Fig.~\ref{fig:perio}, middle)
was then computed on the residuals, still reajusting the offsets,
but keeping the planetary and noise parameters fixed.
A second significant peak is visible at 53.7~d ($\mathrm{FAP} = 7.7\times 10^{-12}$).
As for the first planet, we performed a global fit of all the parameters
after including this planet in the model.
These fitted parameters are presented in Table~\ref{tab:params},
the kernel function corresponding to the fitted hyperparameters is illustrated
in Fig.~\ref{fig:GP_kernel},
and the residuals of each time series superimposed with the GP prediction
are shown in Figs.~\ref{fig:residuals} and \ref{fig:residuals_zoom}.
Finally, the periodogram of the residuals after fitting both planets
(Fig.~\ref{fig:perio}, bottom)
does not show any significant peak ($\mathrm{FAP}$ above 1\%).
Our conclusions, based on this periodogram and FAP approach,
thus agree with the findings of \citet{ahrer_2021_harps},
with two confirmed planets at 14.19~d and 53.7~d.

We note that the first harmonics part ($k_{\mathrm{SHO,\,harm.}}$)
in the MEP kernel (Eq.~(\ref{eq:MEPkernel}))
plays a key role in the GP modeling,
even if its amplitude is significantly smaller than the fundamental term ($k_{\mathrm{SHO,\,fund.}}$).
Indeed, when neglecting the harmonics term (see Fig.~\ref{fig:perio_noharm}),
we find a third highly significant signal around 19~d ($\mathrm{FAP} = 4.3\times 10^{-5}$),
which corresponds to half the period of the GP ($P_\mathrm{GP} \approx 38$~d, see Table~\ref{tab:params}).
On the contrary, when including the harmonics part,
the peak around 19~d is no more significant (see Fig.~\ref{fig:perio}).
This is not surprising since stellar activity is expected to introduce
signals in the RV and indicators time series
at the rotation period,
but also at its harmonics,
and, in particular, the first harmonics.

By design, a kernel which has power at the rotation harmonics will have a tendency
to absorb signals at this period (here 19~d),
even if those are not due to stellar activity.
To further test whether the 19~d signal could be due to a planet,
\citet{hara_2021_testing} checked the consistency of this signal
over the timespan of the data set.
The authors showed that the 19~d signal, while not statistically significant,
exhibits a stable presence accross the data set,
which might motivate further observations.

\subsection{Bayesian framework and false inclusion probability}
\label{sec:fip}

In order to more directly compare our results with the study of \citet{ahrer_2021_harps},
we performed a full Bayesian evidence analysis using
the nested sampling algorithm \textsc{PolyChord} \citep{handley_2015_polychord}.
Recently, \textsc{PolyChord} was used for radial velocity exoplanet detection by
\citet{ahrer_2021_harps}, \citet{rajpaul_2021_harpsn}, and Unger et al. (2021, accepted).
We aimed at reproducing a similar analysis as \citet{ahrer_2021_harps},
but using \spleaftwo{} and the GP model detailed in Sect.~\ref{sec:kernel}.

The prior distributions we used for all parameters are detailed in Table \ref{tab:polychord_priors}.
We kept mostly the same priors used by \citet{ahrer_2021_harps}, with only a few changes.
We set the amplitude for the RV term of the GP to be strictly positive
to avoid degeneracies in the sign of the amplitudes.
We also shifted the prior for the mean longitudes from $[0, 2\pi]$ to $[-\pi, \pi]$.
Indeed, the mean longitude of planet c is very close to 0,
and this shift avoids splitting the peak in the posterior,
thus improving the convergence efficiency.
For all runs we used a number of live points equal to $50\, n_{\rm dim}$,
that is, 50 times the number of free parameters.
For the precision criterion (stopping criterion), we used $10^{-9}$.

\begin{figure}
  \centering
  \includegraphics[width=\linewidth]{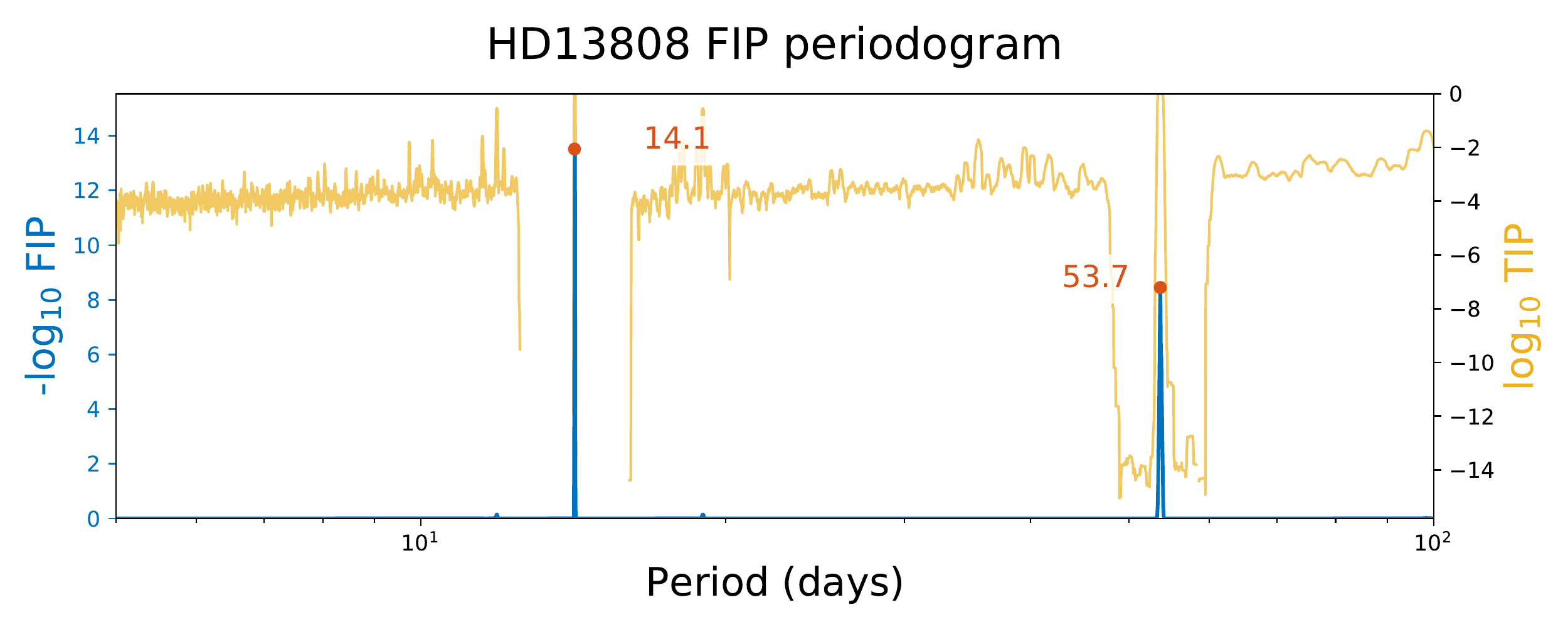}
  \caption{FIP periodogram for HD13808.
    In blue we represent the FIP (false inclusion probability)
    and in yellow the TIP (true inclusion probability).}
  \label{fig:FIP}
\end{figure}

We ran \textsc{PolyChord} for models with zero up to three planets
and each model was run five times to obtain estimates
of the value and uncertainty of the evidence.
The evidence for each model is presented in Table~\ref{tab:evidence}.
We see a steady and significant increase in evidence up to the model with two planets.
The two and three planet models present similar evidence ($\Delta \ln\,Z = 3.5 \pm 2.5$, compatible at 1.4~$\sigma$).
Moreover, even if the three planet model is marginally favored,
no clear period for a third planet emerges from the posteriors.
To illustrate this, we computed the false inclusion probability (FIP) periodogram \citep{hara_2021_improving}
from the posteriors of all our runs with zero to three planets (see Fig.~\ref{fig:FIP}).
The true inclusion probability (TIP) provides the probability
for the system to host at least one planet in a given (small) range of period.
On the contrary, the FIP (=1-TIP) is the probability that the system does not host any planet in the given range of period.
The FIP periodogram of Fig.~\ref{fig:FIP} is computed with a window size of
$1/(t_\mathrm{max}-t_\mathrm{min})$ in frequency.
We observe in Fig.~\ref{fig:FIP} two very significant peaks (FIP less than $10^{-8}$) around 14.1~d and 53.7~d,
which confirms that these planets should be included in the model.
Then, two smaller peaks around 12~d and 19~d are also visible
but none of them is significant (FIP higher than 70\%).
The posterior distribution of parameters of the two-planet runs
is given in Table~\ref{tab:params}.
Our results are in agreement with the periodogram and FAP approach detailed in Sect.~\ref{sec:periofap},
and with \citet{ahrer_2021_harps} findings.

\begin{table}
  \caption{Prior distributions used for each parameter in the nested sampling runs with \textsc{PolyChord}.}
  \centering
  \setlength{\tabcolsep}{5pt}
  \begin{tabular}{ccccc}
    \hline
    \hline
    Parameter                      & (units)                        & Prior             & Lower bound                   & Upper bound             \\ \hline
    $P_\mathrm{GP}$                & (d)                            & $\mathcal{U}$     & 10                            & 100                     \\
    $\rho_\mathrm{GP}$             & (d)                            & $\log\mathcal{U}$ & 10                            & 400                     \\
    $\eta_\mathrm{GP}$             &                                & $\log\mathcal{U}$ & 0.01                          & 10                      \\ \hline
    $\gamma_\mathrm{RV}$           & ($\mathrm{m} \mathrm{s}^{-1}$) & $\mathcal{U}$     & RV$_{\min}$                   & RV$_{\max}$             \\
    $\sigma_\mathrm{RV}$           & ($\mathrm{m} \mathrm{s}^{-1}$) & $\mathcal{U}$     & 0                             & 20                      \\
    $\alpha_\mathrm{RV}$           & ($\mathrm{m} \mathrm{s}^{-1}$) & $\mathcal{U}$     & 0                             & rms(RV)                 \\
    $\beta_\mathrm{RV}$            &                                & $\mathcal{U}$     & $-10\,\mathrm{rms(RV)}$       & $10\,\mathrm{rms(RV)}$  \\ \hline
    $\gamma_\mathrm{BIS}$          & ($\mathrm{m} \mathrm{s}^{-1}$) & $\mathcal{U}$     & $\mathrm{BIS}_{\min}$         & BIS$_{\max}$            \\
    $\sigma_\mathrm{BIS}$          & ($\mathrm{m} \mathrm{s}^{-1}$) & $\mathcal{U}$     & 0                             & rms(BIS)                \\
    $\alpha_\mathrm{BIS}$          & ($\mathrm{m} \mathrm{s}^{-1}$) & $\mathcal{U}$     & $-\mathrm{rms(BIS)}$          & rms(BIS)                \\
    $\beta_\mathrm{BIS}$           &                                & $\mathcal{U}$     & $-10\,\mathrm{rms(BIS)}$      & $10\,\mathrm{rms(BIS)}$ \\ \hline
    $\gamma_{\log R_{HK}'}$        &                                & $\mathcal{U}$     & $\log R_{HK\, \min}'$         & $\log R_{HK\, \max}'$   \\
    $\sigma_\mathrm{\log R_{HK}'}$ &                                & $\mathcal{U}$     & 0                             & rms($\log R_{HK}'$)     \\
    $\alpha_{\log R_{HK}'}$        &                                & $\mathcal{U}$     & $-\mathrm{rms}(\log R_{HK}')$ & rms($\log R_{HK}'$)     \\ \hline
    $P$                            & (d)                            & $\log\mathcal{U}$ & 5                             & 100                     \\
    $\lambda_\mathrm{0}$           & (deg)                          & $\mathcal{U}$     & $-\pi$                        & $\pi$                   \\
    $K$                            & ($\mathrm{m} \mathrm{s}^{-1}$) & $\log\mathcal{U}$ & 0.1                           & 10                      \\
    $e$                            &                                & $\mathcal{U}$     & 0                             & 1                       \\
    $\omega$                       & (deg)                          & $\mathcal{U}$     & 0                             & $2\pi$                  \\ \hline
  \end{tabular}
  \tablefoot{$\mathcal{U}$ stands for uniform distribution,
    while $\log\mathcal{U}$ stands for log-uniform distribution.}
  \label{tab:polychord_priors}
\end{table}

\begin{table}
  \caption{Evidence of each considered model in our \textsc{PolyChord} runs.}
  \centering
  \begin{tabular}{ccc}
    \hline
    \hline
    Model     & $\log Z$           & $\log Z_i - \log Z_2$ \\ \hline
    0 planet  & $-557.23 \pm 0.17$ & $-88.73 \pm 1.4$      \\
    1 planet  & $-497.3 \pm 1.1$   & $-28.8 \pm 1.8$       \\
    2 planets & $-468.5 \pm 1.4$   & $0.0$                 \\
    3 planets & $-465.0 \pm 2.1$   & $+3.5 \pm 2.5$        \\ \hline
  \end{tabular}
  \tablefoot{Uncertainties are calculated as the standard deviation
    of five identical nested sampling runs for each model.}
  \label{tab:evidence}
\end{table}

\section{Conclusion}
\label{sec:conclusion}

In this article, we have presented \spleaftwo{},
a GP framework that is able to efficiently model multiple time series simultaneously.
Classical GP models have a computational cost that scales as the cube of the number of measurements,
which makes them prohibitive in terms of computational resources for large data sets.
The computational cost of our GP framework scales linearly with the data set size,
which allows for tractable computations even for large data sets.

This work builds on previous studies that provided efficient GP models for single time series
\citep[in particular the \celerite{} and \spleaf{} models, see][]{rybicki_1995_class,ambikasaran_2015_generalized,foreman-mackey_2017_fast,delisle_2020_efficientb}.
It is also inspired by a recent generalization of \celerite{} to the case of two-dimensional data sets
\citep{gordon_2020_fast}, but extends it by accounting for the GP derivatives.
These derivatives are especially important when modeling the effect of stellar activity on RV time series
\citep[see][]{aigrain_2012_simple,rajpaul_2015_gaussian}.
Our framework additionally accounts for time series that do not share the same calendar,
which is useful to train a GP simultaneously on RV and photometric measurements taken with two
different instruments \citep[e.g.,][]{haywood_2014_planets}.

We applied our methods to reanalyze the RV time series of the nearby K2 dwarf HD~13808.
Our results are very similar to a recent state-of-the-art study of the same system \citep{ahrer_2021_harps}
and we confirm the two planets announced in this article.
However, we have shown that using our framework allowed us to dramatically decrease (by more than two orders of magnitude)
the computational cost of the GP modeling.
While the data set analyzed here consists of 738 measurements (RV, BIS, and $\rhk$ at 246 epochs),
the gain of using \spleaftwo{} would be even greater for larger data sets.
Some data sets (e.g., HARPS or HARPS-N Sun-as-a-star RV time series) that could not have been analyzed
with such a GP modeling are now achievable with our GP framework.

Finally, it is worth noting that
the results from the periodogram and FAP approach are consistent with
the much more computer intensive Bayesian evidence calculations using nested sampling.
This illustrates the power of the periodogram and FAP computation
including a correlated noise model, as proposed by \citet{delisle_2020_efficient}.

\begin{acknowledgements}
  We thank the referee, D. Foreman-Mackey,
  for his very constructive feedback that helped to improve this manuscript.
  We acknowledge financial support from the Swiss National Science Foundation (SNSF).
  This work has, in part, been carried out within the framework of
  the National Centre for Competence in Research PlanetS
  supported by SNSF.
\end{acknowledgements}

\bibliographystyle{aa}
\bibliography{spleaf2}

\begin{thebibliography}{28}
\expandafter\ifx\csname natexlab\endcsname\relax\def\natexlab#1{#1}\fi

\bibitem[{{Ahrer} {et~al.}(2021){Ahrer}, {Queloz}, {Rajpaul}, {S{\'e}gransan},
  {Bouchy}, {Hall}, {Handley}, {Lovis}, {Mayor}, {Mortier}, {Pepe}, {Thompson},
  {Udry}, \& {Unger}}]{ahrer_2021_harps}
{Ahrer}, E., {Queloz}, D., {Rajpaul}, V.~M., {et~al.} 2021, \mnras, 503, 1248

\bibitem[{{Aigrain} {et~al.}(2012){Aigrain}, {Pont}, \&
  {Zucker}}]{aigrain_2012_simple}
{Aigrain}, S., {Pont}, F., \& {Zucker}, S. 2012, \mnras, 419, 3147

\bibitem[{Ambikasaran(2015)}]{ambikasaran_2015_generalized}
Ambikasaran, S. 2015, Numerical Linear Algebra with Applications, 22, 1102

\bibitem[{{Baluev}(2008)}]{baluev_2008_assessing}
{Baluev}, R.~V. 2008, \mnras, 385, 1279

\bibitem[{{Blackman} {et~al.}(2020){Blackman}, {Fischer}, {Jurgenson},
  {Sawyer}, {McCracken}, {Szymkowiak}, {Petersburg}, {Ong}, {Brewer}, {Zhao},
  {Leet}, {Buchhave}, {Tronsgaard}, {Llama}, {Sawyer}, {Davis}, {Cabot},
  {Shao}, {Trahan}, {Nemati}, {Genoni}, {Pariani}, {Riva}, {Fournier}, \&
  {Pawluczyk}}]{blackman_2020_performance}
{Blackman}, R.~T., {Fischer}, D.~A., {Jurgenson}, C.~A., {et~al.} 2020, \aj,
  159, 238

\bibitem[{{David} {et~al.}(2019){David}, {Petigura}, {Luger}, {Foreman-Mackey},
  {Livingston}, {Mamajek}, \& {Hillenbrand}}]{david_2019_four}
{David}, T.~J., {Petigura}, E.~A., {Luger}, R., {et~al.} 2019, \apjl, 885, L12

\bibitem[{{Delisle} {et~al.}(2020{\natexlab{a}}){Delisle}, {Hara}, \&
  {S{\'e}gransan}}]{delisle_2020_efficient}
{Delisle}, J.~B., {Hara}, N., \& {S{\'e}gransan}, D. 2020{\natexlab{a}}, \aap,
  635, A83

\bibitem[{{Delisle} {et~al.}(2020{\natexlab{b}}){Delisle}, {Hara}, \&
  {S{\'e}gransan}}]{delisle_2020_efficientb}
{Delisle}, J.~B., {Hara}, N., \& {S{\'e}gransan}, D. 2020{\natexlab{b}}, \aap,
  638, A95

\bibitem[{{Dumusque} {et~al.}(2012){Dumusque}, {Pepe}, {Lovis},
  {S{\'e}gransan}, {Sahlmann}, {Benz}, {Bouchy}, {Mayor}, {Queloz}, {Santos},
  \& {Udry}}]{dumusque_2012_earthmass}
{Dumusque}, X., {Pepe}, F., {Lovis}, C., {et~al.} 2012, \nat, 491, 207

\bibitem[{{Dumusque} {et~al.}(2011){Dumusque}, {Udry}, {Lovis}, {Santos}, \&
  {Monteiro}}]{dumusque_2011_planetary}
{Dumusque}, X., {Udry}, S., {Lovis}, C., {Santos}, N.~C., \& {Monteiro},
  M.~J.~P.~F.~G. 2011, \aap, 525, A140

\bibitem[{{Foreman-Mackey}(2018)}]{foreman-mackey_2018_scalable}
{Foreman-Mackey}, D. 2018, Research Notes of the American Astronomical Society,
  2, 31

\bibitem[{{Foreman-Mackey} {et~al.}(2017){Foreman-Mackey}, {Agol},
  {Ambikasaran}, \& {Angus}}]{foreman-mackey_2017_fast}
{Foreman-Mackey}, D., {Agol}, E., {Ambikasaran}, S., \& {Angus}, R. 2017, \aj,
  154, 220

\bibitem[{{Gillen} {et~al.}(2020){Gillen}, {Briegal}, {Hodgkin},
  {Foreman-Mackey}, {Van Leeuwen}, {Jackman}, {McCormac}, {West}, {Queloz},
  {Bayliss}, {Goad}, {Watson}, {Wheatley}, {Belardi}, {Burleigh}, {Casewell},
  {Jenkins}, {Raynard}, {Smith}, {Tilbrook}, \& {Vines}}]{gillen_2020_ngts}
{Gillen}, E., {Briegal}, J.~T., {Hodgkin}, S.~T., {et~al.} 2020, \mnras, 492,
  1008

\bibitem[{{Gordon} {et~al.}(2020){Gordon}, {Agol}, \&
  {Foreman-Mackey}}]{gordon_2020_fast}
{Gordon}, T.~A., {Agol}, E., \& {Foreman-Mackey}, D. 2020, \aj, 160, 240

\bibitem[{{Handley} {et~al.}(2015){Handley}, {Hobson}, \&
  {Lasenby}}]{handley_2015_polychord}
{Handley}, W.~J., {Hobson}, M.~P., \& {Lasenby}, A.~N. 2015, \mnras, 453, 4384

\bibitem[{{Hara} {et~al.}(2021{\natexlab{a}}){Hara}, {Delisle}, {Unger}, \&
  {Dumusque}}]{hara_2021_testing}
{Hara}, N.~C., {Delisle}, J.-B., {Unger}, N., \& {Dumusque}, X.
  2021{\natexlab{a}}, arXiv e-prints, arXiv:2106.01365

\bibitem[{{Hara} {et~al.}(2021{\natexlab{b}}){Hara}, {Unger}, {Delisle},
  {D{\'\i}az}, \& {S{\'e}gransan}}]{hara_2021_improving}
{Hara}, N.~C., {Unger}, N., {Delisle}, J.-B., {D{\'\i}az}, R., \&
  {S{\'e}gransan}, D. 2021{\natexlab{b}}, arXiv e-prints, arXiv:2105.06995

\bibitem[{{Haywood} {et~al.}(2014){Haywood}, {Collier Cameron}, {Queloz},
  {Barros}, {Deleuil}, {Fares}, {Gillon}, {Lanza}, {Lovis}, {Moutou}, {Pepe},
  {Pollacco}, {Santerne}, {S{\'e}gransan}, \& {Unruh}}]{haywood_2014_planets}
{Haywood}, R.~D., {Collier Cameron}, A., {Queloz}, D., {et~al.} 2014, \mnras,
  443, 2517

\bibitem[{{Jones} {et~al.}(2017){Jones}, {Stenning}, {Ford}, {Wolpert},
  {Loredo}, {Gilbertson}, \& {Dumusque}}]{jones_2017_improving}
{Jones}, D.~E., {Stenning}, D.~C., {Ford}, E.~B., {et~al.} 2017, arXiv
  e-prints, arXiv:1711.01318

\bibitem[{{Jord{\'a}n} {et~al.}(2021){Jord{\'a}n}, {Eyheramendy}, \&
  {Buchner}}]{jordan_2021_statespace}
{Jord{\'a}n}, A., {Eyheramendy}, S., \& {Buchner}, J. 2021, Research Notes of
  the American Astronomical Society, 5, 107

\bibitem[{{Mayor} {et~al.}(2011){Mayor}, {Marmier}, {Lovis}, {Udry},
  {S{\'e}gransan}, {Pepe}, {Benz}, {Bertaux}, {Bouchy}, {Dumusque}, {Lo Curto},
  {Mordasini}, {Queloz}, \& {Santos}}]{mayor_2011_harps}
{Mayor}, M., {Marmier}, M., {Lovis}, C., {et~al.} 2011, ArXiv e-prints
  [\eprint[arXiv]{1109.2497}]

\bibitem[{{Noyes} {et~al.}(1984){Noyes}, {Hartmann}, {Baliunas}, {Duncan}, \&
  {Vaughan}}]{noyes_1984_rotation}
{Noyes}, R.~W., {Hartmann}, L.~W., {Baliunas}, S.~L., {Duncan}, D.~K., \&
  {Vaughan}, A.~H. 1984, \apj, 279, 763

\bibitem[{{Pepe} {et~al.}(2021){Pepe}, {Cristiani}, {Rebolo}, {Santos},
  {Dekker}, {Cabral}, {Di Marcantonio}, {Figueira}, {Lo Curto}, {Lovis},
  {Mayor}, {M{\'e}gevand}, {Molaro}, {Riva}, {Zapatero Osorio}, {Amate},
  {Manescau}, {Pasquini}, {Zerbi}, {Adibekyan}, {Abreu}, {Affolter}, {Alibert},
  {Aliverti}, {Allart}, {Allende Prieto}, {{\'A}lvarez}, {Alves}, {Avila},
  {Baldini}, {Bandy}, {Barros}, {Benz}, {Bianco}, {Borsa}, {Bourrier},
  {Bouchy}, {Broeg}, {Calderone}, {Cirami}, {Coelho}, {Conconi}, {Coretti},
  {Cumani}, {Cupani}, {D'Odorico}, {Damasso}, {Deiries}, {Delabre},
  {Demangeon}, {Dumusque}, {Ehrenreich}, {Faria}, {Fragoso}, {Genolet},
  {Genoni}, {G{\'e}nova Santos}, {Gonz{\'a}lez Hern{\'a}ndez}, {Hughes},
  {Iwert}, {Kerber}, {Knudstrup}, {Landoni}, {Lavie}, {Lillo-Box}, {Lizon},
  {Maire}, {Martins}, {Mehner}, {Micela}, {Modigliani}, {Monteiro}, {Monteiro},
  {Moschetti}, {Murphy}, {Nunes}, {Oggioni}, {Oliveira}, {Oshagh}, {Pall{\'e}},
  {Pariani}, {Poretti}, {Rasilla}, {Rebord{\~a}o}, {Redaelli}, {Santana
  Tschudi}, {Santin}, {Santos}, {S{\'e}gransan}, {Schmidt}, {Segovia},
  {Sosnowska}, {Sozzetti}, {Sousa}, {Span{\`o}}, {Su{\'a}rez Mascare{\~n}o},
  {Tabernero}, {Tenegi}, {Udry}, \& {Zanutta}}]{pepe_2021_espresso}
{Pepe}, F., {Cristiani}, S., {Rebolo}, R., {et~al.} 2021, \aap, 645, A96

\bibitem[{{Queloz} {et~al.}(2001){Queloz}, {Henry}, {Sivan}, {Baliunas},
  {Beuzit}, {Donahue}, {Mayor}, {Naef}, {Perrier}, \&
  {Udry}}]{queloz_2001_planet}
{Queloz}, D., {Henry}, G.~W., {Sivan}, J.~P., {et~al.} 2001, \aap, 379, 279

\bibitem[{{Rajpaul} {et~al.}(2015){Rajpaul}, {Aigrain}, {Osborne}, {Reece}, \&
  {Roberts}}]{rajpaul_2015_gaussian}
{Rajpaul}, V., {Aigrain}, S., {Osborne}, M.~A., {Reece}, S., \& {Roberts}, S.
  2015, \mnras, 452, 2269

\bibitem[{{Rajpaul} {et~al.}(2021){Rajpaul}, {Buchhave}, {Lacedelli}, {Rice},
  {Mortier}, {Malavolta}, {Aigrain}, {Borsato}, {Mayo}, {Charbonneau},
  {Damasso}, {Dumusque}, {Ghedina}, {Latham}, {L{\'o}pez-Morales},
  {Magazz{\`u}}, {Micela}, {Molinari}, {Pepe}, {Piotto}, {Poretti}, {Rowther},
  {Sozzetti}, {Udry}, \& {Watson}}]{rajpaul_2021_harpsn}
{Rajpaul}, V.~M., {Buchhave}, L.~A., {Lacedelli}, G., {et~al.} 2021, \mnras,
  507, 1847

\bibitem[{{Rasmussen} \& {Williams}(2006)}]{rasmussen_2006_gaussian}
{Rasmussen}, C.~E. \& {Williams}, C. K.~I. 2006, {Gaussian Processes for
  Machine Learning}

\bibitem[{{Rybicki} \& {Press}(1995)}]{rybicki_1995_class}
{Rybicki}, G.~B. \& {Press}, W.~H. 1995, Physical Review Letters, 74, 1060

\end{thebibliography}

\appendix
\section{Semiseparable representation of Matérn 3/2 and Matérn 5/2 kernels}
\label{sec:matern}

The Matérn 3/2 and 5/2 covariances are widely used in various fields of statistics.
Their kernel functions are written as
\begin{align}
   & k_{3/2}(\Delta t) = \sigma^2 \l(1 + \Delta x\r)\expo{-\Delta x},\nonumber                \\
   & k_{5/2}(\Delta t) = \sigma^2 \l(1 + \Delta x + \frac{1}{3}\Delta x^2\r)\expo{-\Delta x},
\end{align}
where $x$ is the rescaled time:
\begin{align}
  \label{eq:xdef}
   & x = \sqrt{3} \frac{t}{\rho}\qquad \text{for the Matérn 3/2 kernel},\nonumber \\
   & x = \sqrt{5} \frac{t}{\rho}\qquad \text{for the Matérn 5/2 kernel},
\end{align}
and $\sigma$ and $\rho$ are the GP hyperparameters.

\subsection{Semiseparable representation}
\label{sec:matern_semisep}

For two times $t_i > t_j$,
we have
\begin{equation}
  k_{3/2}(t_i-t_j) = \sigma^2 \l(\l(x_i \expo{-x_i}\r) \expo{x_j} + \expo{-x_i}\l((1-x_j)\expo{x_j}\r)\r),
\end{equation}
which provides the semiseparable representation of rank 2
\begin{align}
   & A_i = \sigma^2,\nonumber                   \\
   & U_{i,1} = \sigma^2 x_i \expo{-x_i}, \qquad
  V_{i,1} = \expo{x_i},\nonumber                \\
   & U_{i,2} = \sigma^2 \expo{-x_i},\qquad
  V_{i,2} = (1 - x_i) \expo{x_i}
\end{align}
for the Matérn 3/2 kernel.
Similarly, the Matérn 5/2 kernel admits the semiseparable representation of rank 3
\begin{align}
   & A_i = \sigma^2,\nonumber                                           \\
   & U_{i,1} = \sigma^2 \l(x_i + \frac{x_i^2}{3}\r) \expo{-x_i}, \qquad
  V_{i,1} = \expo{x_i},\nonumber                                        \\
   & U_{i,2} = \sigma^2 \expo{-x_i},\qquad
  V_{i,2} = \l(1 - x_i + \frac{x_i^2}{3}\r) \expo{x_i},\nonumber        \\
   & U_{i,3} = \sigma^2 x_i \expo{-x_i},\qquad
  V_{i,3} = -\frac{2}{3} x_i \expo{x_i}.
\end{align}
These representations are not unique and the choice of splitting into
the 2 (Matérn 3/2) or 3 (Matérn 5/2) semiseparable terms is arbitrary.

\subsection{Derivative of a Matérn Gaussian process}
\label{sec:matern_deriv}

A GP following the Matérn 3/2 or 5/2 kernel is always differentiable, independently of the hyperparameters.
Following the same reasoning as in Sect.~\ref{sec:derivative},
we compute the derivatives of $U$ and $V$,
as well as the matrix $B$,
which appear in the covariance matrix between the GP $G(t)$ and its derivative $G'(t)$
and in the covariance matrix of $G'(t)$ itself
(see Eq.~(\ref{eq:DD2})).
We find
\begin{align}
  U'_{i,1} & = \frac{\sqrt{3}\sigma^2}{\rho} (1 - x_i) \expo{-x_i},\nonumber \\
  U'_{i,2} & = -\frac{\sqrt{3}\sigma^2}{\rho} \expo{-x_i},\nonumber          \\
  V'_{i,1} & = \frac{\sqrt{3}}{\rho} \expo{x_i},\nonumber                    \\
  V'_{i,2} & = -\frac{\sqrt{3}}{\rho} x_i \expo{x_i},\nonumber               \\
  B_i      & = \frac{3 \sigma^2}{\rho^2},
\end{align}
for the Matérn 3/2 kernel,
and
\begin{align}
  U'_{i,1} & = \frac{\sqrt{5}\sigma^2}{3\rho} \l(3-x_i-x_i^2\r) \expo{-x_i},\nonumber \\
  U'_{i,2} & = -\frac{\sqrt{5}\sigma^2}{\rho} \expo{-x_i},\nonumber                   \\
  U'_{i,3} & = \frac{\sqrt{5}\sigma^2}{\rho} (1 - x_i)\expo{-x_i},\nonumber           \\
  V'_{i,1} & = \frac{\sqrt{5}}{\rho} \expo{x_i},\nonumber                             \\
  V'_{i,2} & = \frac{\sqrt{5}}{3\rho} \l(x_i^2-x_i\r)\expo{x_i},\nonumber             \\
  V'_{i,3} & = -\frac{2\sqrt{5}}{3\rho} (1 + x_i)\expo{x_i},\nonumber                 \\
  B_i      & = \frac{5 \sigma^2}{3\rho^2},
\end{align}
for the Matérn 5/2 kernel.

\subsection{Overflows and preconditioning}
\label{sec:matern_overflows}

To avoid overflows (see Sect.~\ref{sec:overflows}),
these representations can be adapted to use the same preconditioning method
as in the case of the classical \celerite{} quasiperiodic terms.
The preconditioning matrix $\phi$ is defined as
\begin{equation}
  \phi_{i,s} = \expo{-(x_{i+1}-x_i)},
\end{equation}
and the preconditioned matrices $\tilde{U}$, $\tilde{V}$, $\tilde{U}'$, and $\tilde{V}'$
are obtained by dropping the exponential terms from the definitions of $U$, $V$, $U'$, and $V'$.
For instance, we have
\begin{align}
   & \tilde{U}_{i,1} = \sigma^2 x_i, \qquad
  \tilde{V}_{i,1} = 1,\nonumber                                         \\
   & \tilde{U}_{i,2} = \sigma^2,\qquad
  \tilde{V}_{i,2} = 1 - x_i,\nonumber                                   \\
   & \tilde{U}'_{i,1} = \frac{\sqrt{3}\sigma^2}{\rho} (1 - x_i), \qquad
  \tilde{V}'_{i,1} = \frac{\sqrt{3}}{\rho},\nonumber                    \\
   & \tilde{U}'_{i,2} = -\frac{\sqrt{3}\sigma^2}{\rho}, \qquad
  \tilde{V}'_{i,2} = -\frac{\sqrt{3}}{\rho} x_i,\nonumber               \\
   & \phi_{i,1} = \phi_{i,2} = \expo{-(x_{i+1}-x_i)},
\end{align}
for the Matérn 3/2 covariance matrix.

While this preconditioning allows us to prevent overflows and underflows due to
the exponential terms,
weaker numerical instabilities could arise due to the presence of
$x_i$ and $x_i^2$ in the definitions of the preconditioned matrices.
The absolute values of the rescaled times $x_i$
should thus be kept as small as possible
to improve numerical stability.
Since the Matérn kernels are stationary (i.e., $k$ only depends on $\Delta t$),
a reference time $t_0$ can be chosen arbitrarily,
and the definition of $x$ (Eq.~(\ref{eq:xdef})) can be adapted
accordingly
\begin{align}
   & x = \sqrt{3} \frac{t-t_0}{\rho}\qquad \text{for the Matérn 3/2 kernel},\nonumber \\
   & x = \sqrt{5} \frac{t-t_0}{\rho}\qquad \text{for the Matérn 5/2 kernel}.
\end{align}
For instance, we could use
\begin{equation}
  t_0 = \frac{\min(t)+\max(t)}{2}
\end{equation}
to avoid $x_i$ values that are too large.
This might not be sufficient for a very large time span
compared to the decay timescale, $\rho$;
particularly in the case of the Matérn 5/2 kernel, which contains
quadratic terms ($x_i^2$).
Recently, \citet{jordan_2021_statespace} proposed a state-space representation
for the Matérn 3/2 and 5/2 kernels, which allows a similar linear scaling
of the likelihood evaluation with improved numerical stability.

\section{Twice mean square differentiable semiseparable kernels}
\label{sec:twicediff}

In this Appendix we construct several
twice mean square differentiable semiseparable kernels.
When modeling time series as combinations of a GP and its derivative,
the chosen GP kernel must at least be once mean square differentiable
for the model to be valid.
However, using a twice mean square differentiable kernel
ensures that the GP's derivative is itself differentiable,
which typically generates smoother models.

\subsection{Alternatives to the SE kernel}

\begin{figure}
  \centering
  \includegraphics[width=\linewidth]{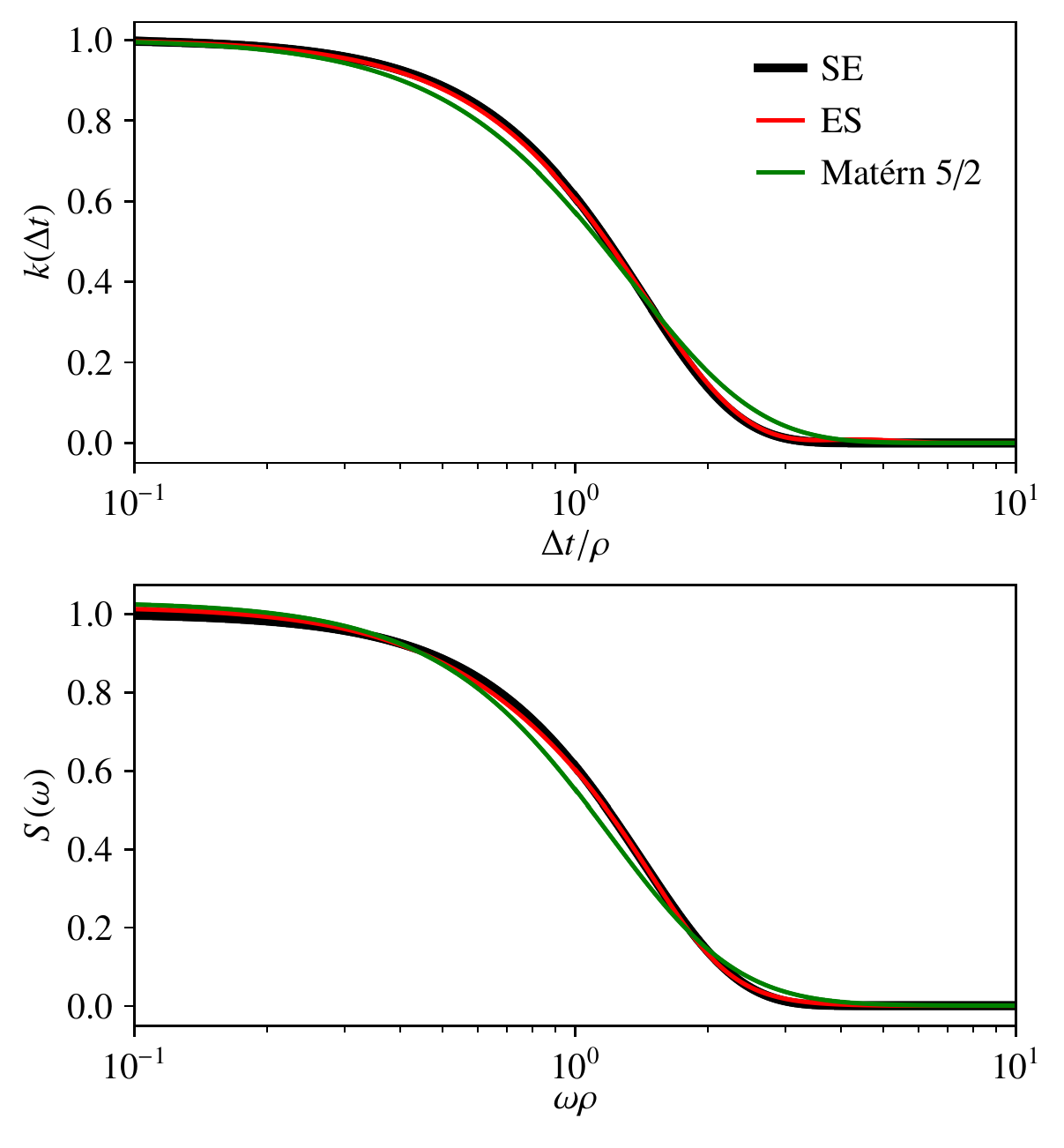}
  \caption{Comparison of the kernel functions (\textit{top}) and
    power spectral density (\textit{bottom}) of the SE, ES, and Matérn 5/2 kernels.
    The timescale of the Matérn 5/2 kernel is adjusted such as to minimize the maximum deviation from the SE kernel.
  }
  \label{fig:SEkernel_comp}
\end{figure}

The Matérn 5/2 kernel is a widely spread kernel which has the advantage of
being both twice differentiable and semiseparable with rank $r=3$
(see Appendix~\ref{sec:matern}).
If twice differentiability is required,
it is thus a natural alternative to the squared-exponential (SE) kernel:
\begin{equation}
  k_\mathrm{SE}(\Delta t) = \sigma^2\exp\l(-\frac{\Delta t^2}{2\rho^2}\r)
\end{equation}
since the latter cannot be modeled with \celerite{}/\spleaf{}.
Higher order Matérn kernels, such as the Matérn 7/2 kernel could also be used
as they are more than twice differentiable and admit semiseparable representations.
However, the rank of their semiseparable representations would be higher,
which would increase the cost of likelihood evaluations.

We additionally propose here the exponential-sine (ES) kernel
\begin{align}
  k_\mathrm{ES}(\Delta t) = \sigma^2 \expo{-\lambda \Delta t}
   & \l(1 + \frac{1-2\mu^{-2}}{3} \l(\cos(\mu\lambda\Delta t) - 1\r)\r.\nonumber \\
   & \l.\vphantom{\frac{1-2\mu^{-2}}{3}}+ \mu^{-1}\sin(\mu\lambda\Delta t)\r),
\end{align}
which is also twice differentiable and semiseparable with rank 3.
Its cost is thus similar to the Matérn 5/2 kernel.
The corresponding power spectral density (PSD)
\begin{equation}
  S_\mathrm{ES}(\omega) = \frac{2\sqrt{2}}{3\sqrt{\pi}} \frac{\sigma^2\l(1 + \mu^2\r) \l(4 + \mu^2\r)}{\lambda \l(1+\l(\omega/\lambda\r)^2\r) \l(\l(1+\l(\omega/\lambda\r)^2 - \mu^2\r)^2 + 4\mu^2\r)}
\end{equation}
is always positive (for $\lambda>0$),
which ensures the positive definiteness of the kernel
\citep[e.g.,][]{foreman-mackey_2017_fast}.
The parameters $\lambda$ and $\mu$ can be chosen arbitrarily, but
using
\begin{align}
  \lambda & \approx \frac{1.091}{\rho},\nonumber \\
  \mu     & \approx 1.327,
\end{align}
makes the deviation between the SE and the ES kernels
below $0.009\sigma^2$ for all lags $\Delta t$.
Figure~\ref{fig:SEkernel_comp} illustrates this by comparing
the SE, ES, and Matérn 5/2 kernels and PSD.

\subsection{Alternatives to the SEP kernel}

\begin{figure}
  \centering
  \includegraphics[width=\linewidth]{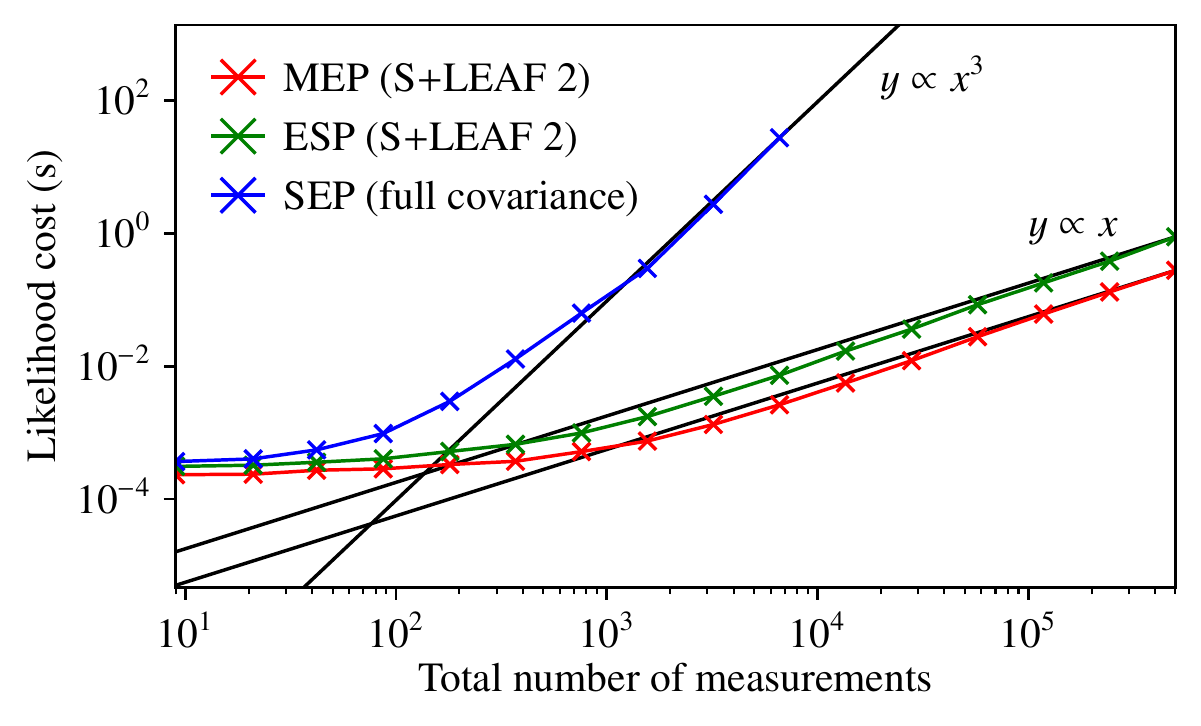}
  \caption{Same as Fig.~\ref{fig:perfs} but including the ESP kernel in the
    performance comparison.}
  \label{fig:perfs_esp}
\end{figure}

\begin{figure}
  \centering
  \includegraphics[width=\linewidth]{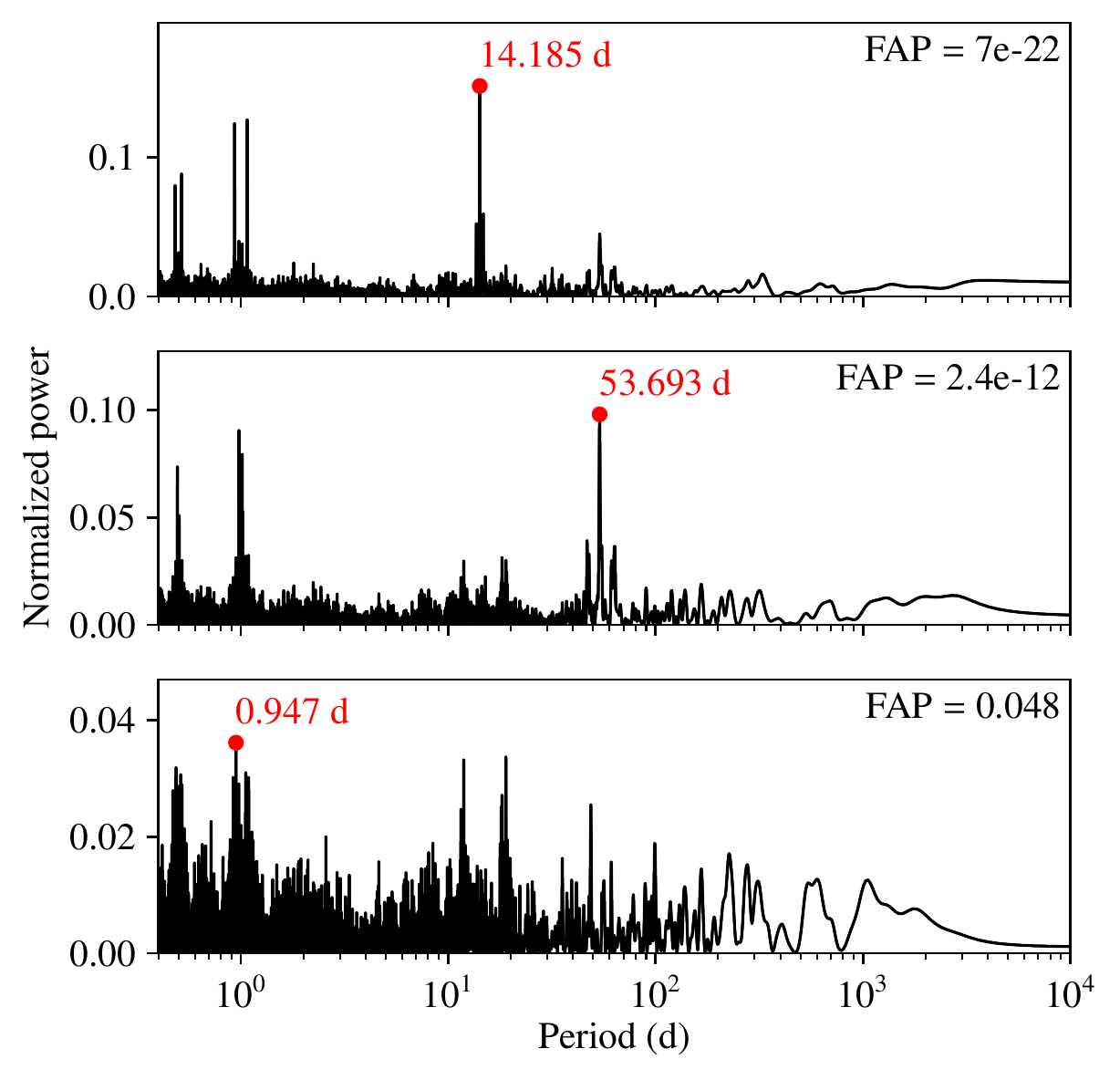}
  \caption{Same as Fig.~\ref{fig:perio} but using the ESP kernel instead of the MEP kernel.}
  \label{fig:perio_esp}
\end{figure}

\begin{figure}
  \centering
  \includegraphics[width=\linewidth]{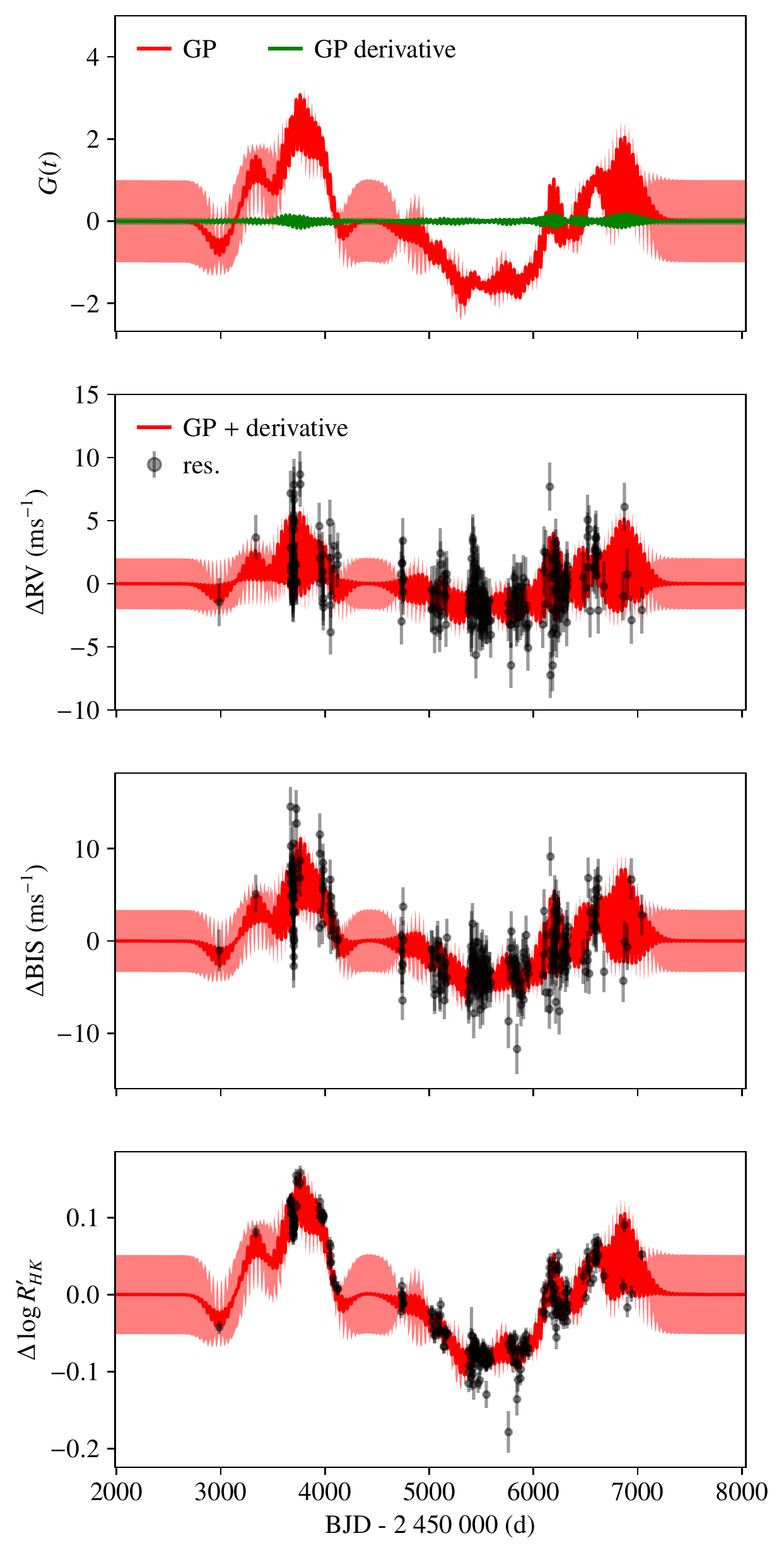}
  \caption{Same as Fig.~\ref{fig:residuals} but using the ESP kernel instead of the MEP kernel.}
  \label{fig:residuals_esp}
\end{figure}

\begin{figure}
  \centering
  \includegraphics[width=\linewidth]{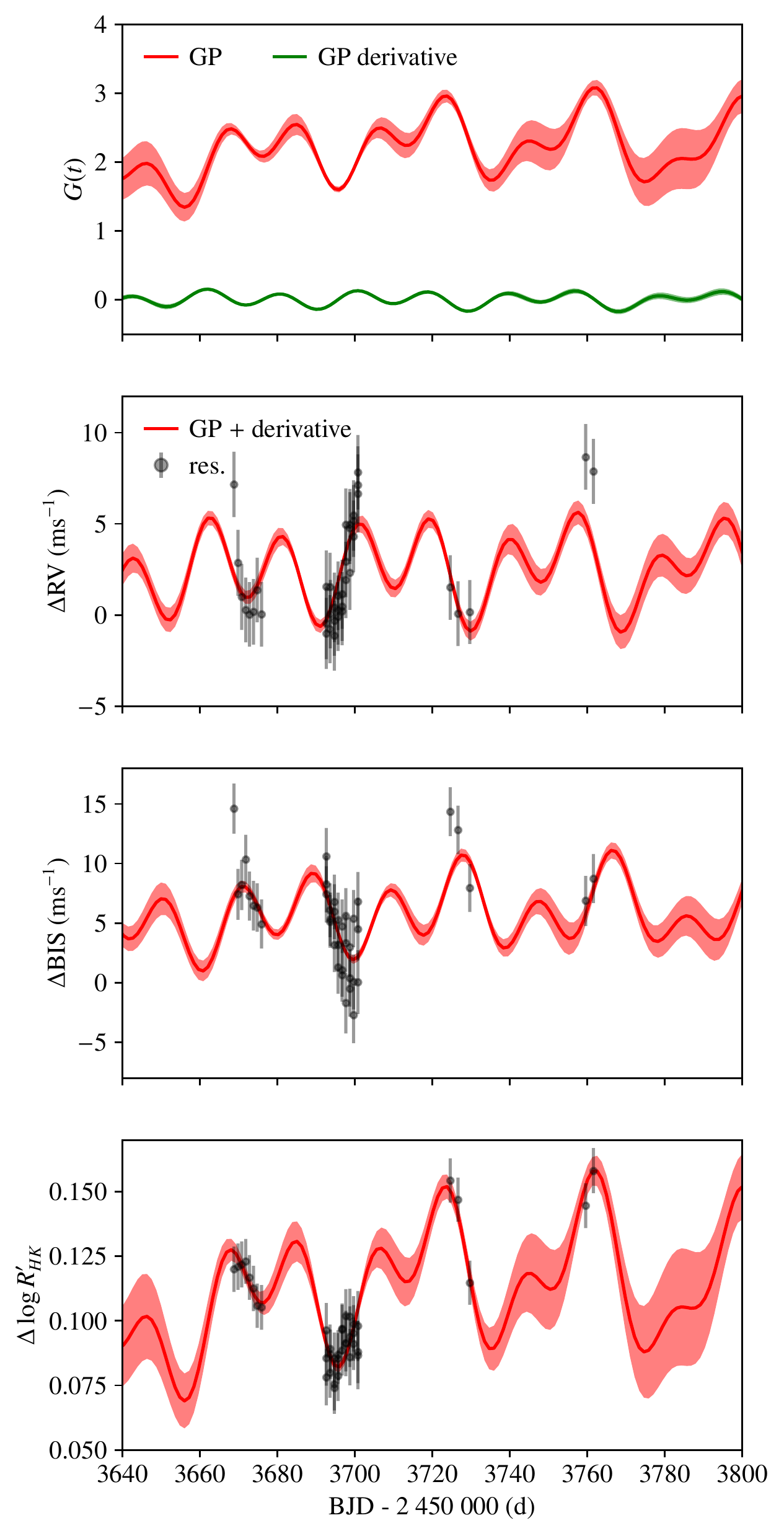}
  \caption{Same as Fig.~\ref{fig:residuals_zoom} but using the ESP kernel instead of the MEP kernel.}
  \label{fig:residuals_zoom_esp}
\end{figure}

\begin{figure}
  \centering
  \includegraphics[width=\linewidth]{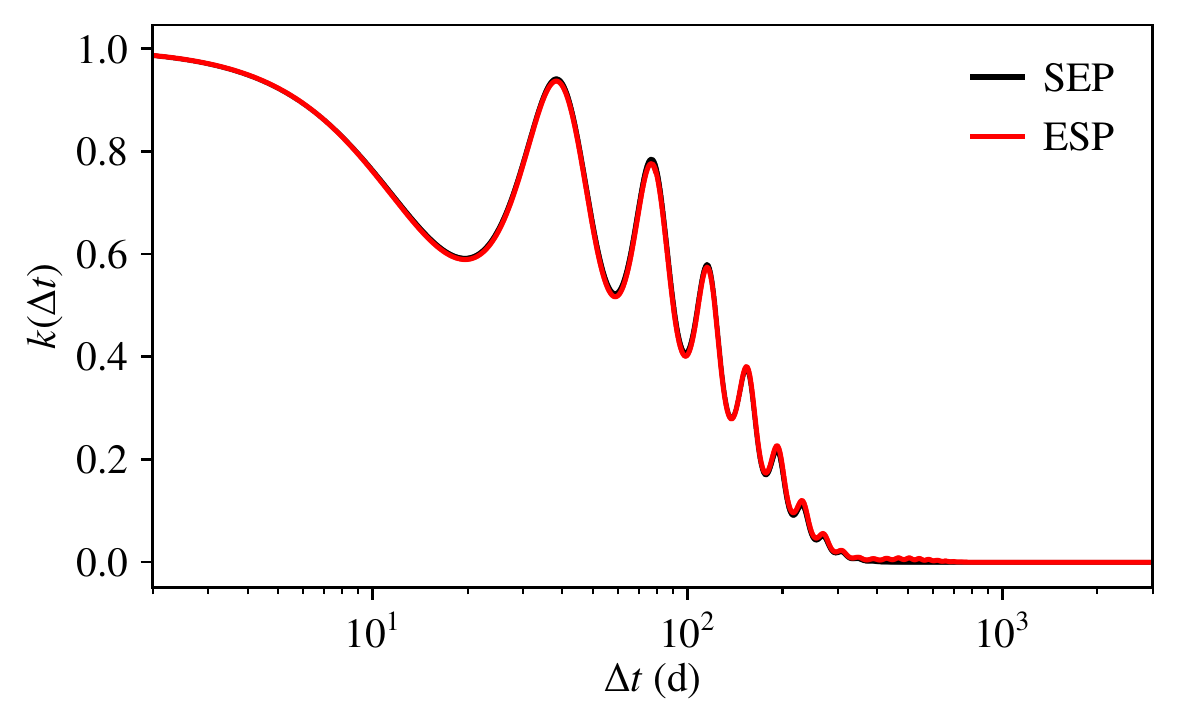}
  \caption{Same as Fig.~\ref{fig:GP_kernel} but using the ESP kernel instead of the MEP kernel.}
  \label{fig:GP_kernel_esp}
\end{figure}

As seen in Eq.~(\ref{eq:SEPdev}) the SEP kernel can be approximated by
\begin{equation}
  \label{eq:SEPdevb}
  k_\mathrm{SEP}(\Delta t) \approx k_\mathrm{SE}(\Delta t)
  \frac{1 + f\cos\l(\nu \Delta t\r)
    + \frac{f^2}{4} \cos\l(2\nu\Delta t\r)}{1+f+\frac{f^2}{4}}.
\end{equation}
In this expression, the periodic part
\begin{equation}
  k_\mathrm{P}(\Delta t) = \frac{1 + f\cos\l(\nu \Delta t\r)
    + \frac{f^2}{4} \cos\l(2\nu\Delta t\r)}{1+f+\frac{f^2}{4}}
\end{equation}
is semiseparable and verifies $k'_\mathrm{P}(0)=k^{(3)}_\mathrm{P}(0)=0$.
Thus, in order to obtain a twice differentiable semiseparable kernel
similar to the SEP kernel,
one simply needs to replace the SE part in Eq.~(\ref{eq:SEPdevb})
by a Matérn 5/2 or ES kernel and we define
\begin{align}
  k_\mathrm{5/2 P}(\Delta t) & = k_{5/2}(\Delta t) k_\mathrm{P}(\Delta t),       \\
  k_\mathrm{ESP}(\Delta t)   & = k_\mathrm{ES}(\Delta t) k_\mathrm{P}(\Delta t).
\end{align}
Indeed, the product of two semiseparable terms is semiseparable
\citep[see][]{foreman-mackey_2017_fast}
and since
\begin{align}
  (fg)'      & = f'g + fg',\nonumber                   \\
  (fg)^{(3)} & = f^{(3)}g + 3f''g' + 3f'g'' + fg^{(3)}
\end{align}
the first and third derivatives of $k_\mathrm{5/2 P}$ and $k_\mathrm{ESP}$
also cancel out at $\Delta t = 0$.
The PSD of these two kernels are given by
\begin{align}
  S_{k\mathrm{P}}(\omega) = \frac{1}{1 + f + \frac{f^2}{4}} & \l(
  S_k(\omega)
  + f\frac{S_k(\omega+\nu)+S_k(\omega-\nu)}{2}\r.\nonumber                                                                  \\
                                                            & \l.+ \frac{f^2}{4}\frac{S_k(\omega+2\nu)+S_k(\omega-2\nu)}{2}
  \r),
\end{align}
where $k=5/2$ or ES.
Since the PSD ($S_k$) of the Matérn~5/2 and ES kernels are strictly positive for all frequencies
and the coefficient $f=(2\eta)^{-2}$ is strictly positive,
we find that $S_{5/2\mathrm{P}}$ and $S_\mathrm{ESP}$
are also strictly positive for all frequencies.
The two corresponding kernels are thus positive definite.

The rank of the semiseparable representations of
$k_{5/2\mathrm{P}}$ and $k_\mathrm{ESP}$ is $r=15$,
since they are the product of a rank 3 kernel (Matérn 5/2 or ES)
and a rank 5 kernel (periodic part).
As a comparison, the MEP kernel (see Eq.~\ref{eq:MEPkernel}),
which is not twice differentiable,
is of a rank of 6.

We reproduced the analyses of Sects.~\ref{sec:perfs} and \ref{sec:periofap}
using the ESP kernel instead of the MEP kernel.
The results are presented in Figs.~\ref{fig:perfs_esp}-\ref{fig:GP_kernel_esp}.
The cost of likelihood evaluations using the ESP kernel
is about twice the cost of using the MEP kernel (see Fig.~\ref{fig:perfs_esp}),
which is still much more efficient than modeling the full covariance matrix.
The periodograms (Fig.~\ref{fig:perio_esp}),
as well as the GP prediction (Figs.~\ref{fig:residuals_esp} and \ref{fig:residuals_zoom_esp})
are very similar to the ones obtained with the MEP kernel (Figs.~\ref{fig:perio},~\ref{fig:residuals}, and \ref{fig:residuals_zoom}).
Finally, we see in Fig.~\ref{fig:GP_kernel_esp} that the ESP kernel
reproduces the SEP kernel very closely while the MEP kernel
mimics it more roughly (see Fig.~\ref{fig:GP_kernel}).
However, these differences between the MEP and the ESP (or SEP) kernels
seem to have a very weak impact on our analysis,
since the periodograms and GP prediction are similar in both cases.

\section{Periodogram and FAP for heterogeneous time series}
\label{sec:fapmodif}

We consider here the case of an heterogeneous time series
following Eq.~(\ref{eq:RajpaulGeneral})
and we are aimed at detecting a periodic signal affecting the first time series $Y_{1}$ only.
The frameworks of \citet{baluev_2008_assessing} and \citet{delisle_2020_efficient}
defining a general class of linear periodograms
with their associated analytical FAP approximations
can be applied to the merged time series $y$ of Eq.~(\ref{eq:RajpaulFlat}) with a slight modification.
We thus refer to \citet{delisle_2020_efficient} for the details of the framework
and we focus here on the required adaptations.
Following \citet{delisle_2020_efficient}, we compare the $\chi^2$ of a linear base model $\H$
of $p$ parameters with enlarged models $\K$ of $p+d$ parameters, parameterized by the frequency $\nu$.
The base model is defined as
\begin{equation}
  \H\ :\quad m_\H(\theta_\H) = \varphi_\H\theta_\H,
  \label{eq:mh}
\end{equation}
where $\theta_\H$ is the vector of size $p$ of the model parameters,
$\varphi_\H$ is a $n\times p$ matrix,
and $n$ the total number of points in the merged time series $y$.
The columns of $\varphi_H$ are explanatory time series
that are scaled by the linear parameters $\theta_\H$.
For instance, if we consider the merged time series of RV, BIS, and $\rhk$,
and we include in the model an offset for each of these time series,
we would have to define:
\begin{equation}
  m_\H = \gamma_\mathrm{RV} \delta_\mathrm{RV} + \gamma_\mathrm{BIS} \delta_\mathrm{BIS} + \gamma_{\rhk} \delta_{\rhk},
\end{equation}
where $\gamma_i$ is the offset of time series $i$,
and $\delta_i$ is equal to one for measurements belonging to time series $i$ and zero otherwise.
The matrix $\varphi_\H$ would thus be a $n\times 3$ matrix,
whose first column would be filled with 1 for RV measurements and 0 otherwise,
the second column would be equal to 1 for BIS measurements,
and the last column for $\rhk$ measurements.
The vector of parameters would then be $\theta_\H = (\gamma_\mathrm{RV}, \gamma_\mathrm{BIS}, \gamma_{\rhk})$.

The enlarged model $\K(\nu)$ is written as
\begin{equation}
  \K(\nu)\ :\quad m_\K(\nu, \theta_\K) = \varphi_\K(\nu)\theta_\K,
  \label{eq:mk}
\end{equation}
where $\theta_\K = (\theta_\H, \theta)$ is the vector of size $p+d$ of the parameters
and $\varphi_\K(\nu) = (\varphi_H, \varphi(\nu))$ is a $n\times (p+d)$ matrix
whose $p$ first columns are those of $\varphi_\H$,
and whose $d$ last columns are functions of the frequency, $\nu$.
In the case of an homogeneous time series, as in \citet{delisle_2020_efficient},
one typically uses $\varphi(\nu) = (\cos(\nu t),\ \sin(\nu t))$ (with $d=2$).
The main difference in the case of heterogeneous time series
is that we only search for a periodicity in the first time series ($Y_1$, typically the RV time series).
Thus, we would define
$\varphi(\nu) = (\cos(\nu t) \delta_1,\ \sin(\nu t) \delta_1)$.
All the results presented in \citet{delisle_2020_efficient} remain valid
when applied to the merged time series.
The only difference is that due to the presence of zeroes in $\varphi$ for all measurements
not belonging to the first time series,
the averaging used in the definition of the effective time series length (\citet{delisle_2020_efficient} Eq.~(8))
is restricted to the first time series measurements
\begin{equation}
  \l<X\r> = \sum_{\substack{i,j,\\\delta_1(i) = \delta_1(j) =1}} C_{i,j}^{-1} X_{i,j},
\end{equation}
where $C^{-1}$ is the inverse of the full covariance matrix of the merged time series.

\end{document}